\documentclass[prb,twocolumn,superscriptaddress]{revtex4}
\usepackage[T1]{fontenc}
\usepackage{color}
\usepackage{graphicx}
\usepackage{amsmath}
\usepackage{amssymb}
\usepackage{amsfonts}
\usepackage[colorlinks,hyperindex]{hyperref}
\usepackage{epsfig}

\def\beq{\begin{equation}}
\def\eeq{\end{equation}}

\begin{document}

\title{Neuromorphic computing in Ginzburg-Landau polariton lattice systems}
\author{Andrzej Opala}
\affiliation{Institute of Physics, Polish Academy of Sciences, Al. Lotnik\'ow 32/46,PL-02-668 Warsaw, Poland}
\email{opala@ifpan.edu.pl}
\author{Sanjib Ghosh}
\affiliation{Division of Physics and Applied Physics, Nanyang Technological University 637371, Singapore}
\author{Timothy C. H. Liew}
\affiliation{Division of Physics and Applied Physics, Nanyang Technological University 637371, Singapore}
\author{Micha{\l} Matuszewski}
\affiliation{Institute of Physics, Polish Academy of Sciences, Al. Lotnik\'ow 32/46,PL-02-668 Warsaw, Poland}

\begin{abstract}
The availability of large amounts of data and the necessity to process it
efficiently have led to rapid development of machine learning techniques. To name a
few examples, artificial neural network architectures are commonly used for
financial forecasting, speech and image recognition, robotics, medicine, and even research.
Direct hardware for neural networks is highly sought for overcoming the von Neumann bottleneck of software implementations. Reservoir computing (RC) is a recent and increasingly popular
bio-inspired computing scheme which holds promise for an efficient temporal
information processing. We demonstrate the applicability and performance of
reservoir computing in a general complex Ginzburg-Landau lattice model, which adequately
describes dynamics of a wide class of systems, including coherent photonic devices.
In particular, we propose that the concept can be readily applied in
exciton-polariton lattices, which are characterized by unprecedented photonic
nonlinearity, opening the way to signal processing at rates of the order of 1 Tbit
s$^{-1}$. 
\end{abstract}

\maketitle

\section{Introduction}

In contrast to single- or multilayer (deep) feedforward networks, recurrent neural
networks (RNNs) may exhibit complex internal state dynamics. This makes them
particularly efficient in the analysis of time-dependent signals that require
memory, such as speech or text recognition and
interpretation~\cite{LeCun_DeepLearning}.
On the other hand, training of RNNs is difficult and not always convergent.
The idea of reservoir computing (RC), also known as echo state networks or liquid
state machines, may be viewed as a generalization of the RNN concept, inspired by
the internal structure of the
brain~\cite{Jaeger_HarnessingESN,Maass_RealTimeComputing,Enel_RCPropertiesPrefrontalCortex,Lukosevicius_RCapproachestoRNN}.

\begin{figure}
\includegraphics[width=0.5\textwidth]{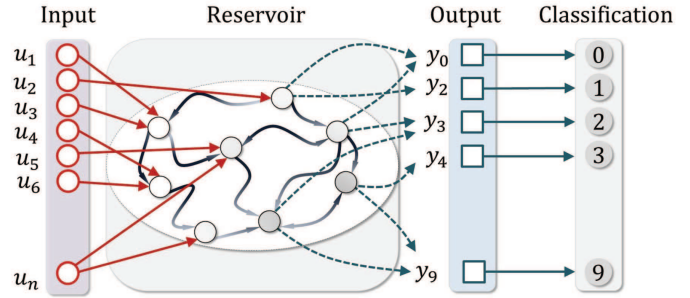}
\caption{Concept of reservoir computing. A single input layer $u_i(t)$ is used to
excite signals in the nonlinear reservoir consisting of hidden nodes. Signals
propagate between the nodes connected with each other with random weights. This
performs a nonlinear transformation of input in a high-dimensional space. The
evolution of node amplitudes is collected in the output layer $y_i$ and used for
classification or prediction. In contrast to standard recurrent neural networks, the
input weights and connections within the reservoir are static. Only the output
weights are trained using a convergent regression procedure.}
\label{fig:Fig_1}
\end{figure}

The core of the system is formed by a recurrent network of nodes (neurons) connected
with each other, see Fig.~\ref{fig:Fig_1}. This network, called the reservoir, is
static as its connections are unchanged during training. This is important for both
the feasibility of a physical implementation and the convergence of the training
procedure. Typically, the input signal represented by $u_i(t)$ is multiplied by
random weights and injected into reservoir nodes. The subsequent dynamics of
amplitudes in the nodes (called neuron activations) is delivered to output neurons
$y_i$ that are used e.g.~for classification or prediction of time-dependent signals.
In the classification task, the output neuron with the highest activation is
identified with the predicted class. A supervised training procedure consists of
presenting many different inputs to the system, and adjusting the output weights to
minimize the average error. Output signals are linear superpositions of the
amplitudes of the reservoir, which results in a training procedure that is both
efficient and convergent~\cite{Lukosevicius_PracticalGuideESN}.

The simple scheme described above proved to be surprisingly efficient in various
machine learning tasks. The performance of RC in time series prediction and speech
recognition is particularly well
described~\cite{Jaeger_HarnessingESN,Larger_HiSpeedReservoirComputing,Brunner_ParallelIPGigabyte,Fischer_InformationProcessingSingleNode,Fujii_HarnessingQuantumDynamicsML,Vandoorne_RCSiliconChip,Torrejon_Neuromorphic}.
Notably, software implementation of RC won a financial time series prediction
competition~\cite{Jaeger_competition}. To date, RC has been successfully realized
not only as a software, but also as hardware implementations in systems including
semiconductor chips~\cite{VLSI_Chip}, memristor
arrays~\cite{Du_MemristorRC,Kudithipudi_NeuromemristiveRCBiosignalProcessing}
optoelectronic~\cite{Paquot_OptoelectronicRC,Larger_BeyondTuring,Larger_HiSpeedReservoirComputing}
and all-optical systems~\cite{Duport_AllOpticalRC}, mechanical
oscillators~\cite{Coulombe_ComputingMechanicalOscillators} and even in a bucket of
water~\cite{Bucket}. See~\cite{Tanaka_review} for a comprehensive review. Photonic systems are particularly interesting as they hold
promise for huge signal processing
rates~\cite{Vandoorne_RCSiliconChip,Brunner_ParallelIPGigabyte,Larger_HiSpeedReservoirComputing,Soljacic_DeepLearning,Caulfield_FutureSupercomputing}.

In many machine learning approaches, input data is transformed in a multidimensional
space according to a nonlinear map. This nonlinear mapping in RC is typically
provided by the complex dynamics of the reservoir, although nonlinear readout has
also been implemented~\cite{Vandoorne_RCSiliconChip}. To be useful for computing,
the network has to posses several  characteristics. First, the size of the reservoir
must be sufficiently large, so that it is able to perform the desired computation
within its fixed internal structure. The so-called echo state
property~\cite{Lukosevicius_RCapproachestoRNN} is related to the stability of the
system. The state of the reservoir must be determined solely by the history of the
signal $u_i$, but for sufficiently long evolution it should not depend on the
initial conditions or signals from the distant past. On the other hand, the dynamics
must be sufficiently nonlinear, so that the state of the reservoir allows for the
separation of signals that differ by a small amount. The optimal working point
appears to be placed close to a certain stability
threshold~\cite{Vandoorne_RCSiliconChip,Lukosevicius_PracticalGuideESN}.
Importantly, it has been demonstrated that powerful computation can be achieved in a
variety of different designs and systems. For instance, according to the original
idea, reservoir nodes are connected with each other with random
weights~\cite{Jaeger_HarnessingESN}. However, various other designs proved to be
efficient as
well~\cite{Torrejon_Neuromorphic,Vandoorne_RCSiliconChip,Larger_HiSpeedReservoirComputing}.

In this work, we consider the implementation of RC in systems described by the complex
Ginzburg-Landau equation (CGLE), which is one of the fundamental models of wave
phenomena~\cite{Aranson_CGLEWorld}. It provides a universal description of weakly
nonlinear spatiotemporal systems invariant under a global gauge change
$\psi\rightarrow\psi {\rm e}^{i \phi}$, where typically $\psi$ is a slowly varying
envelope of an oscillatory wave packet. Its range of applications spans from the
description of hydrodynamic systems and chemical reactions, to superconductors and
superfluids, to ultracold quantum gases and
lasers~\cite{Aranson_CGLEWorld,Berloff_VortexLattices,Ott_BistabilityDrivenDissipative}.
We propose to implement reservoir dynamics in a lattice of weakly coupled traps with
nearest neighbor couplings. According to our estimates, semiconductor exciton-polariton microcavity systems appear to be a promising platform for reservoir computing with very high signal processing data rates, which can be achieved thanks to the extremely strong optical nonlinearity on a picosecond timescale~\cite{Krizhanovskii_UltralowPowerSolitons}.

\section{Model}

\begin{figure*}
\includegraphics[width=1.0\textwidth]{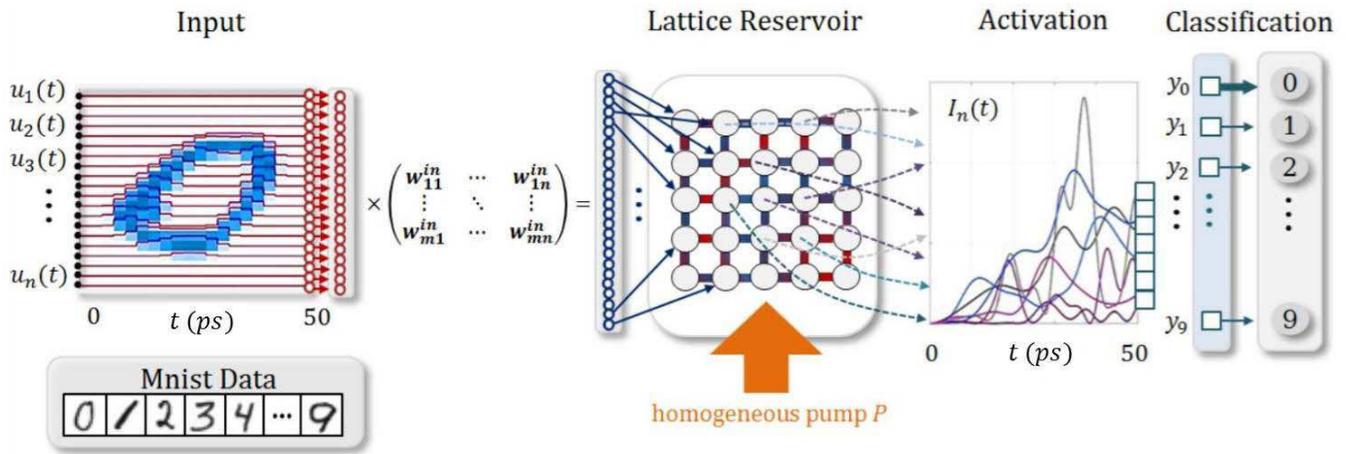}
\caption{Scheme for a handwritten digit classification task. Data is convoluted with
random weights and imprinted on the lattice by driving each of the lattice sites. At
the same time, the system is pumped to maintain a dynamic state close to the
stability (or lasing) threshold. The resulting density $I_n(t)=|\psi_n(t)|^2$ in
each node (activations) is recorded at the end of the sequence and used for
classification of the input.}
\label{fig:Fig_2}
\end{figure*}

We consider the discrete version of the CGLE, which describes a system
enclosed in a simple two-dimensional array of weakly coupled traps
\begin{align}\label{eq:CGLE}
  \frac{\textrm{d}\psi_n}{\textrm{d}t}&=W^{\rm in}_{nm} u_m -i\sum_{m=nn} W_{nm}
\psi_m+\nonumber \\
  &+ \left(\gamma - \Gamma |\psi_n|^2 - i g|\psi_n|^2\right)\psi_n, 
\end{align}
where the first term on the right hand side corresponds to coherent signal injection
with $W^{in}_{nm}$ being the mask applied to the signal $u_m$, see
Fig.~\ref{fig:Fig_2}. Coupling coefficients between nearest-neighbor lattice sites
are denoted by $W_{nm}$, $\gamma=P-\kappa$ is the gain coefficient, in general equal
to the difference between the pumping rate ($P$) and the linear decay rate ($\kappa$), $\Gamma$ is the nonlinear
decay rate, and $g$ is the conservative nonlinear coefficient.
The geometry of the lattice and the scheme of the experiment designed for a
classification task is shown in Fig.~\ref{fig:Fig_2}. In our numerical simulations,
we consider a simple rectangular $N\times N$ lattice reservoir with random positive
symmetric nearest-neighbor couplings, which is a natural choice for photonic systems
such as microcavities, photonic crystals or waveguide arrays. Note that simpler
geometries, such as a one-dimensional lattice with identical couplings and even
single site systems have been demonstrated to perform well in experimental
tests~\cite{Fischer_InformationProcessingSingleNode,Paquot_OptoelectronicRC,Larger_HiSpeedReservoirComputing,Torrejon_Neuromorphic}.
The two-dimensional multisite system has, however, advantage in terms of increased
efficiency.

The parameters of the model~(\ref{eq:CGLE}) used in numerical simulations are chosen
to correspond to a lattice of coupled semiconductor exciton-polariton
microcavities ~\cite{Kim_2013,Winkler_2016,Amo_SOCoupling,Krizhanovskii_LiebLattice}. Exciton-polaritons are composite
quantum quasiparticles of semiconductor excitations and photons in the strong
coupling regime~\cite{Hopfield,Weisbuch,Kavokin}. They exhibit an interesting combination of properties of matter and
light. The extremely low effective mass of polaritons, of the order of $10^{-4}$
electron mass, results from the photonic component, and allows for the effective
transport across the lattice on a picosecond timescale. On the other hand, the
interaction resulting from the exciton component provides unprecedented
instantaneous nonlinearity $g$, orders of magnitude stronger than in other photonic
systems~\cite{Krizhanovskii_UltralowPowerSolitons}. These properties have been used
recently to demonstrate remarkable phenomena including nonequilibrium condensation
and lasing and superfluidity of
polaritons~\cite{Carusotto_QuantumFluids,Kasprzak_BEC,Amo_Superfluidity,Yamamoto_RMP}
as well as realization of ultrafast all-optical
switches~\cite{Sanvitto_Transistor,Savvidis_TransistorSwitch,Bramati_SpinSwitches}.
Recently, a single layer neural network design was proposed~\cite{Liew_Perceptrons}.
Note that reservoir computing should not be confused with the exciton
reservoir, which consists of incoherent particles in the lasing regime. Here, we
neglect the influence of exciton reservoir on the dynamics of the system, which
can be achieved by an appropriate pumping scheme~\cite{Baumberg_SpinBifurcations}.
Our results apply to a range of other systems thanks to the universality of the CGLE
equation and its scaling properties, which allow to convey them to other systems
with different values of physical parameters in the CGLE equation~(\ref{eq:CGLE})
(details of the rescaling are given in the Appendix A).

\section{Results}

We present results for handwritten digit recognition using the modified National
Institute of Standards and Technology (MNIST) dataset, which is one of the standard
tests of pattern recognition in machine learning. Additional simulations for the
Mackey-Glass nonlinear system prediction task and speech recognition are presented in the Appendix D and E. The goal of
the MNIST recognition task is to classify the digits of various writers using the
recorded grayscale images. The dataset contains 70,000 digits, and in our simulations for training (4,000 digits) and testing (1,000 digits) we used randomly picked subsets of the whole set. Each digit consists of
20$\times$20 grayscale pixels.

We convert each image of a digit into temporal signals according to the scheme
illustrated in Fig.~\ref{fig:Fig_2}, with pixels in each row converted into
step-wise signals $u_i(t)$, where $i=1..20$ is the row number. The temporal length
$\tau$ corresponding to a single pixel is adjusted to achieve optimal recognition
rate. This parameter must  be adjusted so that the timescales of the reservior and of
the input signal are compatible, but the overall performance is not very sensitive
to its value. The signal vector $u_i(t)$ is multiplied by a random, constant matrix
${\bf W}^{in}$ of dimension $20^2\times N^2$, which has the purpose of both distributing the information across the lattice and adjusting the length of the input vector to the
number of pillars in the lattice $N^2$. The incoming signal initiates dynamics of
complex reservoir amplitudes $\psi_{n}(t)$, which correspond to neuron activations. The squared modulus of each lattice site is recorded at the end of the sequence. A
linear transformation is used to translate the readout into output neuron
activations, $y_j=\sum_n W^{out}_{jn} |\psi_n(t_E)|^2$, where $j=0..9$ and $t_E=20
\tau$ is the length of the sequence. In the training phase, logistic regression is
used to obtain optimal output weights ${\bf W}^{out}$ (see Appendix C). During testing, the
obtained ${\bf W}^{out}$ are used to classify the digit. Ideally, the result is
equal to $d_j=1$ for the correct digit and $d_j=0$ for all other digits. In
practice, all $d_i$ have mixed values, and we choose the one that is the highest as
the predicted digit.

\begin{figure}
\includegraphics[width=0.4\textwidth]{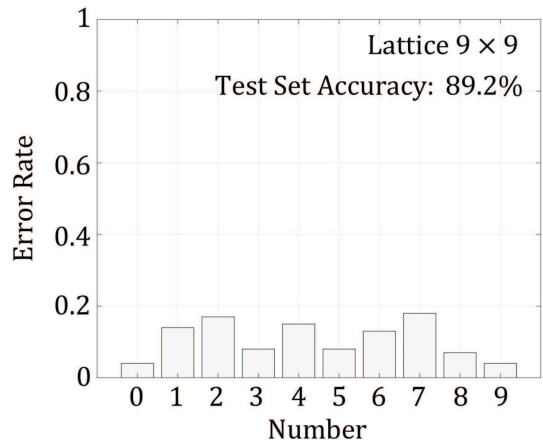}
\caption{Classification error rate for the MNIST dataset for each handwritten digit,
in the case of a $9\times 9$ lattice with 81 nodes.}
\label{fig:Fig_3}
\end{figure}

An example of the resulting error rates is presented in Fig.~\ref{fig:Fig_3}. The
average accuracy for this particular $9\times 9$ lattice is $89.2\%$,
higher than the one obtained by a linear
classifier~\cite{LeCun_MNIST}. Additionally, we note that due to the reduced
dimensionality of the output ($N^2$) with respect to the input ($20^2$), the
regression performed during the learning phase requires much less computation time.
This is important if the minimization of error, performed offline, turns out to be
the most time consuming part of teaching. At the same time, no offline computations
are required during testing, and the recognition rate  is limited only by the
reservoir dynamics timescale.

We used realistic parameters that correspond to experiments performed in gallium
arsenide polariton lattices~\cite{Amo_SOCoupling,Krizhanovskii_LiebLattice}, with
couplings $W_{nm}$ distributed randomly between zero and $0.165\,$meV, $g=0.25 \mu$eV,
and $\Gamma=0.1\mu$eV, $\tau=2.5\,$ps.  Signal processing on a picosecond timescale can be achieved
thanks to the use of a photonic system with a very strong nonlinearity in the 
regime of quantum coupling of light and matter. According to our simulation, the
typical rate at which data can be fed into each lattice site and processed
efficiently is one byte (understood as a unit of information) every few tens of
picoseconds. 
This estimation is in agreement with numerous time-resolved experiments
in polariton systems performed in the nonlinear
regime~\cite{Bobrovska_DI,Deveaud_HydrodynamicVortices,Bloch_MQST,Lagoudakis_Skyrmion,Pieczarka_RelaxationOscillations}.
In contrast to single site RC systems, here signal processing is performed at $N^2$
nodes in a truly parallel manner. A lattice of a hundred of nodes should enable a
realistic data rate of the order of 1~Tbit~s$^{-1}$, in a micrometer sized system.
This compares favorably even with state of the art
optoelectronic~\cite{Brunner_ParallelIPGigabyte} and passive photonic
microcircuit~\cite{Vandoorne_RCSiliconChip} RC implementations, which recently
achieved 10~Gbit~s$^{-1}$ data rates.

\begin{figure}
\includegraphics[width=0.4\textwidth]{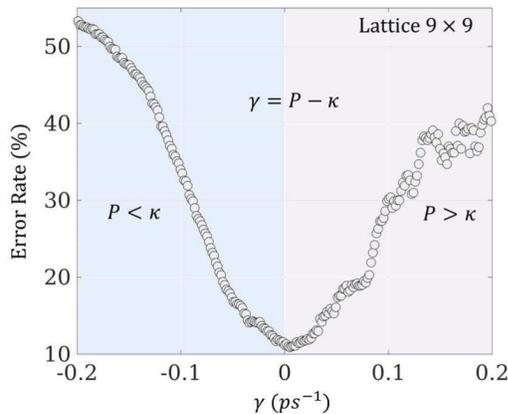}
\caption{Dependence of the error rate on the pumping bias parameter $\gamma$. The
optimal working conditions appears to be close to the stability threshold,
$\gamma\approx 0$.}
\label{fig:Fig_4}
\end{figure}

Figure~\ref{fig:Fig_4} shows the dependence of the error rate on the effective gain
parameter $\gamma$. RC systems display optimal performance when in the absence of
input signal the system is stable, but close to an instability threshold. In the
case of CGLE, at the zero gain point $\gamma=0$ the trivial solution ($\psi_n=0$)
loses stability and a new stationary state appears. This threshold is interpreted in the exciton-polariton context as the onset of polariton lasing. It is clear from
Fig.~\ref{fig:Fig_4} that an optimal working point is found close to zero gain, or
at the lasing threshold.

\begin{figure}
\includegraphics[width=0.4\textwidth]{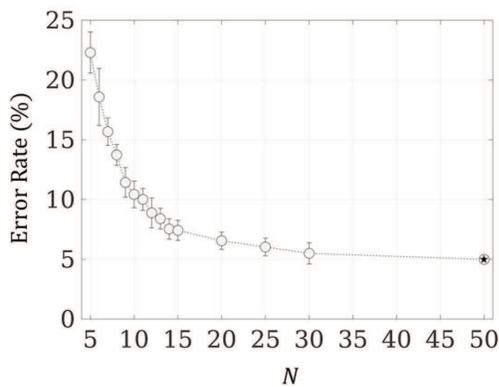}
\caption{Dependence of the error rate on the linear size of the lattice $N$, with
the total number of lattice sites equal to $N^2$. Data points were obtained by
averaging over 10 random realizations of the reservoir weights matrix ${\bf W}$ and
the input matrix ${\bf W^{in}}$. Error bars indicate standard deviation of error rate. Note that the point marked with a star corresponds to a
single realization.}
\label{fig:Fig_5}
\end{figure}

We present the dependence of the error rate on the size of the lattice in
Fig.~\ref{fig:Fig_5}. In this Figure, error rates are calculated as a result of
averaging over 10 different simulations corresponding to different random weights in
the reservoir ${\bf W}$ and in the input matrix ${\bf W^{in}}$. Since the input data
$u_i$ is convoluted with random weights before being used for excitation, adjusting
the number of rows in the rectangular matrix ${\bf W}^{in}$ allows for imprinting
the data on an arbitrary sized lattice. While in the case $N>20$ redundancy in the
input is unavoidable, the same signal may be processed in various ways in different
parts of the reservoir, which leads to improvement of the overall performance.
Indeed, as shown in Fig.~\ref{fig:Fig_5}, the error rate consistently decreases with
$N$, and the error rate for a $50\times50$ lattice is as low as $5.0\%$, similar to
a feedforward neural network with a single hidden layer~\cite{LeCun_MNIST}.
The recognition rate in the case of a $9\times 9$ lattice is equal to $89.2\%$, similar as
in a memristor array RC of comparable size~\cite{Du_MemristorRC}, but the readout vector
size is much smaller in our case (100 variables versus a $176\times 10$ network).

As the error rate decreases with the system size, one can expect that the larger Hilbert space of quantum systems could offer increased performance~\cite{Fujii_HarnessingQuantumDynamicsML} compared to their classical counterparts. Indeed, in the Appendix F, we find a reduced error rate if one has access to additional non-classical observables. However, given the additional complexity of measuring these quantities, the quantum advantage would not necessarily be more practical.

\section{Discussion}

In this work, we demonstrate the applicability of the reservoir computing framework in
the wide class of systems described by the CGLE. 
We would like to point out that the majority of previously considered systems, including optoelectronic ones, were based on amplitude-dependent nonlinearity. On the other hand, the investigated CGLE model describes evolution of a complex wavefunction, where the couplings between nodes are represented by imaginary values, as in the second term in Eq. (1), and the nonlinearity relies on phase modulation (the g interaction constant), which is very different from the amplitude nonlinearity. Our results extend the applicability of reservoir computing to an important class of weakly nonlinear wave systems with gauge invariance, which include coherent photonic systems.
 Importantly, our scheme is robust to both disorder and dissipation, which are usually hindrances in information processing schemes. Here, dissipation is useful for ensuring the echo state property, and a disordered network of connections is part of the design.
We also demonstrate robustness of the model against spatiotemporal noise (see Appendix B).

To illustrate the theoretical performance of the proposed system, we estimate the data processing rate and compare it with hardware implementations of reservoir neural networks  realized experimentally. As the MNIST digit recognition task has not been implemented in many works on RC, we use the TI 46 speech recognition task as a benchmark. The details of the  implementation in the case of an exciton-polariton network are given in Appendix E. Our numerical simulations demonstrate the estimated processing rate of $1.6 \times 10^{10}$ words/s, compared to $2 500$ words/s reported in a CMOS FPGA implementation of liquid state machines~\cite{Wang_LSM_FPGA}, and the record high processing rate of $7.7 \times 10^5$ words/s achieved in an optoelectronic delay line system~\cite{Larger_HiSpeedReservoirComputing}. The very high estimated processing rate of exciton-polariton systems results from the strong optical nonlinearity on a picosecond timescale and the parallel processing in each node of the lattice. 

In terms of scalability of the system size, we note that polariton lattices with several thousands of pillar nodes have already been fabricated and investigated experimentally~\cite{Bloch_EdgeStates}. One can also estimate the physical size scalability with respect to other (non-RC) neuromorphic implementations. For example, the IBM TrueNorth chip~\cite{Merolla_MillionSpkingNeurons} contains one million neurons on an approximate surface of 2 cm$^2$, which gives an average of 5 000 neurons/mm$^2$. A typical size of a polariton pillar node is $10\,\mu$m$^2$, which gives an estimate of 100 000 nodes/mm$^2$. While nodes in the RC framework are not equivalent to neurons in other architectures, which usually have some tunability, this estimate indicates that at least in some applications polariton RC could be competitive with respect to state-of-the-art neuromorphic systems. 

From the point of view of energy efficiency, an important advantage of exciton-polariton systems is that they belong to the class of photonic (neutral particle) systems, which do not suffer from radiative heating, an important issue limiting further development of CMOS and other electronic technologies. Since excitons are neutral particles, they also do not contribute to radiative heating. Energy loss in polariton lattices results mostly from the escape of photons through imperfect microcavity mirrors. However, this does not impose a fundamental limit on energy efficiency (contrary to radiative heating in electronic systems), and several solutions can be proposed to suppress this loss channel. For example, transverse photon modes trapped by total internal reflection can be used to reduce polariton decay through the mirrors~\cite{Krizhanovskii_UltralowPowerSolitons}.

\acknowledgments

AO acknowledges support from the National Science Center, Poland grant No.~2016/22/E/ST3/00045. MM acknowledges support from the National Science Center, Poland grant No.~2017/25/Z/ST3/03032 under QuantERA programme. SG and TL acknowledge support from the Singapore Ministry of Education (via the AcRF Tier 2 grant MOE2017-T2-1-001).
\appendix
\section{Scaling properties of the complex Ginzburg-Landau equation}

The discrete complex Ginzburg--Landau equation with noise reads 
\begin{align}\label{eq:CGLE_app}
  \frac{\textrm{d}\psi_n}{\textrm{d}t}&=W^{\rm in}_{nm} u_m -i\sum_{m=nn} W_{nm}
\psi_m+\nonumber \\
  &+ \left(\gamma - \Gamma |\psi_n|^2 - i g|\psi_n|^2\right)\psi_n + D \xi_n(t), 
\end{align}
The simplicity of Eq.~(\ref{eq:CGLE_app}) allows for the rescaling of physical
coefficients using two arbitrary scaling parameters $\tau$ and $\alpha$, according
to $t=\tau \tilde{t}$, $\psi_n=\alpha^{1/2}\tilde{\psi}_n$,
$u_n=\alpha^{1/2}\tilde{u}_n$, $\gamma=\tilde{\gamma}/\tau$, ${ W}^{in}=
\tilde{ W}^{in}/\tau$, ${ W}_{nm} = \tilde{ W}_{nm}/\tau$,
$g=\tilde{g}/(\tau\alpha)$, $\Gamma= \tilde{\Gamma}/(\tau\alpha)$. The dynamics of
the system with tildes will be identical to the original one except for the
difference in the timescale and amplitude of the wavefunction. It follows that the
only relevant parameters that govern the qualitative behavior of the system are the
ratio $g/\Gamma$, and the relative values of coefficients $\gamma$, $W^{in}_{nm}$, $W_{nm}$ and $\xi_n(t)$. 
In numerical simulations, the random values in the  nearest-neighbor weight matrix ${\bf W}_{nm}$ and in ${\bf W}^{in}$
 matrix are created by the uniform pseudorandom number generator. The function $\xi_n(t)$ is 
 representing the Langevin noise with $D$ being the noise strength.

\section{The effect of noise}
\begin{figure}[h]
\includegraphics[width=0.9\columnwidth]{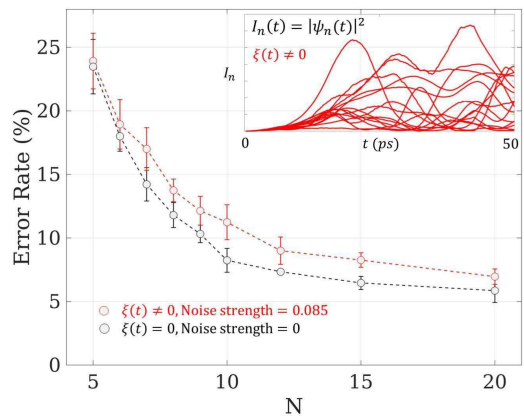}
\caption{Dependence of the error rate on the linear size of the lattice $N$ for 
the MNIST recognizing task, in the cases with and without noise. 
Data is averaged over $5$ random realizations of the reservoir weights 
matrix $\mathbf{W}$, input matrix $\mathbf{W}^\mathrm{in}$ and noise $\xi(t)$. 
The parameters are the same as in Fig.~5 in the main text. The inset plot shows
 the dynamics of density $I_n(t)$ in $n=15$ nodes with nonzero noise.}
\label{fig:MNISTErrorSize}
\end{figure}

The additional Langevin noise term $\xi_n(t)$ is representing noise in the system which is random in space and time.
The effect of noise on the prediction error for the MNIST and speech recognition tasks is presented in  
Figs.~\ref{fig:MNISTErrorSize} and \ref{fig:SpeechErrorSize}, respectively.

\section{Logistic regression}
We use a logistic regression algorithm to train the readout function for the digit recognition task. 
The values between 0 and 1 are assigned by a linear regression classifier to the output 
values for each vector ${\bf y}$ containing $N^2$ elements. 

We introduce the hypothesis function $h_\Theta(y)$ given by
\begin{equation}
h_\Theta({\bf y})={\bf(\Theta^Ty}),
\end{equation}
where ${\bf \Theta}$ is the weights vector.
For  classification of the  hypothesis  representation we introduce the function $g(z)$
\begin{equation}
h_\Theta({\bf y})=g({\Theta^Ty}),
\end{equation}
where $g(z)$ is given by the logistic function
\begin{equation}
g(z)=\frac{1}{{1+e^{-z}}}.
\end{equation}
Combining the above equations the hypothesis function reads 
\begin{equation}
h_\Theta({\bf y})=\frac{1}{{1+e^{-{\bf \Theta^Ty}}}}.
\end{equation}
The cost function is described by the equation
\begin{align}
J(\Theta)&=\frac{1}{n}\sum_{i=1}^{n}\Big[-{\bf x}^{(i)}\log\big(h_\Theta({\bf y}^{(i)})\big)\\
\nonumber&-\big(1-{\bf x^{(i)}}\big)\log\big(1-h_\Theta\big({\bf y}^{(i)}\big)\big) \Big],
\end{align}
where $n$ is the number of samples and $x(i)$ is the correct output for given input states $y(i)$. We minimize the cost function by the gradient method
\begin{equation}
\frac{\partial J({ \bf \Theta})}{\partial { \bf \Theta}_j}=\frac{1}{m}\sum_{i=1}^{n}\big(h_\Theta\big({\bf y}^{(i)}\big)-{\bf x}^{(i)}\big){\bf y}^{(i)}_j.
\end{equation}
Weights are calculated  using Matlab 2016 software with function "fmincg()" written by Carl Edward Rasmussen.

\section{The Mackey-Glass prediction task}

\begin{figure}[h]
\includegraphics[width=1\columnwidth]{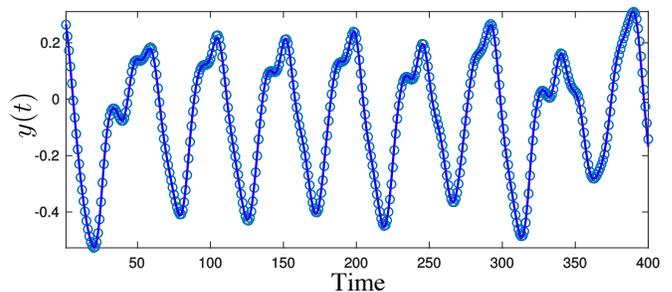}
\caption{The Mackey-Glass prediction task performed by a reservoir computer formed 
with a Ginzburg-Landau lattice. We show $y(t)$ evaluated from Eq.~\ref{Eq:MG} (points) and predicted 
by the reservoir computer (solid line) as functions of time $t$. The overall prediction error $\sigma = 3\times 10^{-4}$ (NRSE value). 
Here we consider $N=16$, $\Gamma=5$, $\gamma =10^{-4}$, $g=2$ and a training dataset of $1000$ time steps.}
\label{fig:MGAccuratePrediction}
\end{figure}

Using the reservoir, we want to predict the solution of the Mackey-Glass equation:
\begin{equation}
 \frac{\partial z}{\partial t} =  \frac{\alpha\,z(t-\tau_{MG})}{1+z^\beta(t-\tau_{MG})}-\gamma_{MG} \, z(t) 
 \label{Eq:MG}
\end{equation}
which is a nonlinear differential equation with time delay feedback. We use the parameters $\alpha = 0.2$, $\gamma_{MG} = 0.1$, $\tau_{MG} = 17$, $\beta = 10$. Using a set of training data we would like to find output weights $W^\text{out}$ such that

\begin{eqnarray}
y(t) = z(t)-1 = \sum_n W^\text{out}_n |x_n(t)|^2
\end{eqnarray}
where the feedback $u=\sum_n W^\text{out}_n |x_n(t-\Delta t)|^2$, where $\Delta t$ is a small time step. 
As a performance measure, we use the normalized mean square error (NRSE):
 \begin{eqnarray}
\sigma = \frac{\sum_i [y(t_i) - y_{p}(t_i)]^2 }{\sum_i [y(t_i) + y_{p}(t_i)]^2 }
\end{eqnarray}
where $y(t_i)$ and $y_{p}(t_i)$ are the true solution of the Mackey-Glass equation and the corresponding prediction from the reservoir computer at time $t_i$, respectively.

In Fig.~\ref{fig:MGAccuratePrediction}, we show the performance of our reservoir computer (a Ginzburg-Landau lattice with $N=16$) for the Mackey-Glass prediction task. For a linear lattice size $N=16$, we found an NRSE $\sigma = 3\times 10^{-4}$ that is similar to the reported values in Ref.~\cite{Li2012}.

\section{Speech recognition} Here we consider the task of isolated spoken digit recognition, which has been a commonly considered benchmarking task for reservoir computing systems~\cite{Paquot_OptoelectronicRC,Larger_BeyondTuring,Vandoorne_RCSiliconChip}. We use a standard data set, which was collected at Texas Instruments (TI) in 1980 (The NIST TI 46 corpus, which is available from the Linguistic Data Consortium). The data is a set of ten isolated spoken digits ($0$ to $9$) by $8$ different female individuals. In the training set, each individual uttered $10$ times a digit, resulting $800$ total spoken digits. After training the network, we evaluate the success rate of our system by using an additional $560$ spoken digits ($10$ digits spoken $7$ times by the $8$ individuals).

\begin{figure}[t]
\includegraphics[width=0.9\columnwidth]{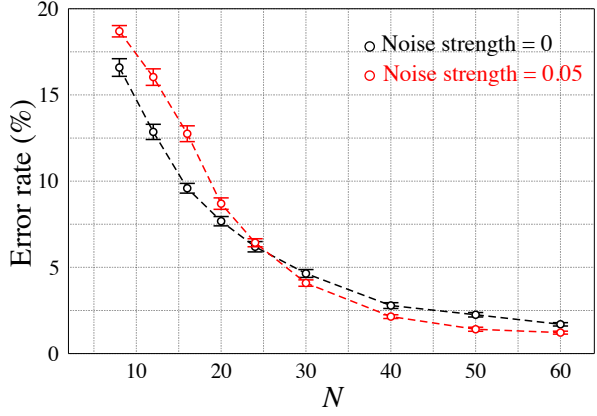}
\caption{The linear lattice size $N$ versus the error rate for isolated spoken digit recognitions with (red) and without (black) noise. We find a noisy system has lower error rate than that of a system without noise when $N$ is large. This could be due to the overfitting of the speech data with a large number of degrees of freedom (for large $N$) which typically diminishes in presence of random noise. The data is averaged over $10$ random realizations of $\mathbf{W}$ and $\mathbf{W}^\mathrm{in}$. The parameters are $\Gamma = 5$, $\gamma=10^{-4}$, $g=2$ and noise strengths $D=0$ (black) and $D=0.05$ (red).}
\label{fig:SpeechErrorSize}
\end{figure}

Each recorded piece of speech is sampled at 12.5 kHz, which is then converted into a cochleagram using the Lyon cochlear ear model~\cite{Lyon1982}, previously identified as a good form of preprocessing for speech recognition~\cite{Verstraeten2005}. These cochleagrams are used as input to the reservoir computer using the same scheme as illustrated in Fig. 2 of the main paper. At a given time $t$, the input temporal signals $u_n(t)$ represent a column of the cochleagram data with $n = 1\dots 78$ (row number). The full cochleagram data sent through $24$ (column number) time steps. The elements of the constant matrix $\textbf{W}^\text{in} $ of size $78 \times N^2$ are chosen random between $ \pm0.5$ ($78\times 24$ is the dimension of the cochleagram data and $N$ is the system size). We readout the computed $|\psi_n|^2$ after sending a full cochleagram signal. The final output is then obtained with the linear transform $y_j = \sum_n W^\text{out}_{jn} |\psi_n|^2$ where $j=0,1\dots 9$. Using the training data set, we obtain the optimal output weights $\textbf{W}^\text{out}$. In the test phase, we use the $560$ test data. The output is recognized as the digit $j$ if $y_{j}$ is the maximum among $j=0,1  \dots 9$. The error rate in recognizing the spoken digits is presented in Fig~\ref{fig:SpeechErrorSize}. As an example for comparison to other works, we obtained a smaller error rates than the reported values for the same in Ref.~\cite{Verstraeten2005}.

\section{An example of computing with quantum reservoirs}

Let us introduce a quantum version of the reservoir network considered in the main article described by the master equation for the system density matrix $\boldsymbol{\rho}$:
\begin{align}
i\hbar\frac{\partial  \boldsymbol{\rho}}{\partial t}&=
i\hbar ( \gamma_Q/2)\sum_n \left( 2 \hat{a}_n \boldsymbol{\rho} \hat{a}_n^\dagger   -\hat{a}_n^\dagger \hat{a}_n \boldsymbol{\rho} - \boldsymbol{\rho} \hat{a}_n^\dagger \hat{a}_n \right)\notag\\
&\hspace{10mm}+\left[ \hat{H}, \boldsymbol{\rho} \right]
\end{align}
where the Hamiltonian is given by,
\begin{align}
\hat{H}&= \sum_{\langle n,m \rangle} J_{nm} \left( \hat{a}_n^\dagger \hat{a}_m +\hat{a}_m^\dagger \hat{a}_n \right) + U \sum_n \hat{a}_n^\dagger a_n^\dagger \hat{a}_n \hat{a}_n\notag\\
&\hspace{10mm}+ u \sum_n \left( F_n \hat{a}_n^\dagger +F_n^* \hat{a}_n \right)
\label{Eq:Hamiltonian}
\end{align}
where the hopping amplitudes $J_{nm}$ and pump strengths $F_n$ are chosen randomly. In a straight forward analogy with the classical version we may define
\begin{align}
y^\text{out}(t)&= \sum_n W^\text{out}_n \text{Tr} [\boldsymbol{\rho}(t) \hat{a}_n^\dagger \hat{a}_n ]\notag\\
&\equiv  \sum_n W^\text{out}_n \langle \hat{a}_n^\dagger(t) \hat{a}_n(t)\rangle
\end{align}
and the feedback $u=y^\text{out}(t-\Delta t) $. However, such a setting has no quantum advantage, as the number of outputs $y^\mathrm{out}(t)$ remains the same and $\langle \hat{a}_n^\dagger(t) \hat{a}_n(t)\rangle$ is an effectively classical quantity, not representing itself any quantum correlations.
\begin{figure}[h]
\includegraphics[width=0.95\columnwidth]{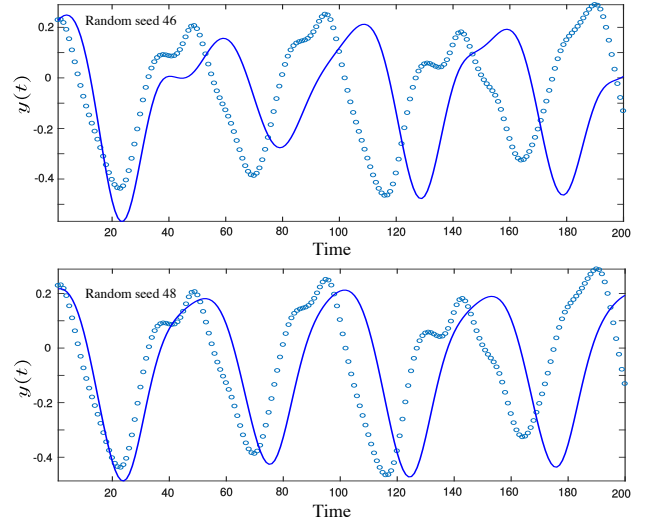}
\caption{Typical prediction (solid lines) from classical reservoir of size $2\times 3$ compared with the Mackey-Glass data (circles). The error $\sigma \sim 0.5$ (NRSE value). We use $\Gamma = 5,\, \gamma =10^{-4},\, g=2 $ and a training dataset of $200$ time steps.}
\label{fig:ClassicalPrediction}
\end{figure}
\begin{figure}[h]
\includegraphics[width=0.95\columnwidth]{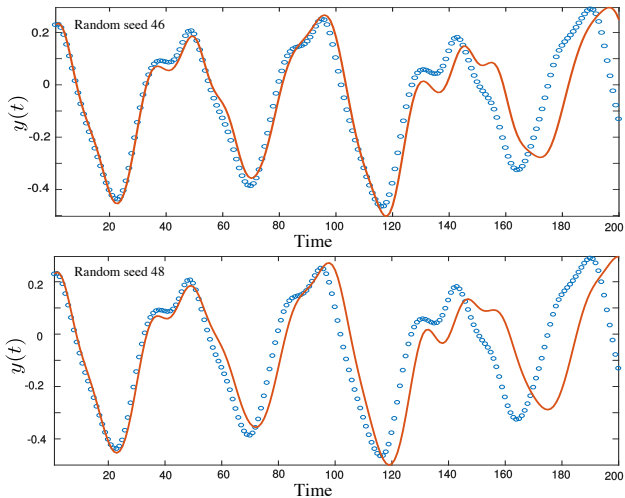}
\caption{Typical prediction (solid lines) from quantum reservoir of size $2\times 3$ compared with the Mackey-Glass data (circles). The error $\sigma \sim 0.01$ (NRSE value). We use $ U=5\gamma_Q$, random $J$ uniformly distributed in $[\pm\gamma_Q/2]$ and a training dataset of $200$ time steps. We see that a quantum reservoir has more prediction capability than that of a classical one with same size.}
\label{fig:QuantumPrediction}
\end{figure}
\begin{figure}[h]
\includegraphics[width=0.95\columnwidth]{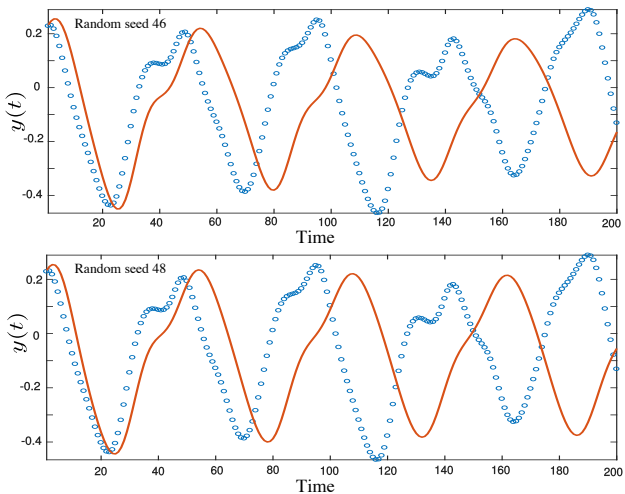}
\caption{Predictions (solid lines) from a highly dissipative quantum reservoir of size $2\times 3$ compared with the Mackey-Glass data (circles). The error $\sigma \sim 0.51$ (NRSE value) , similar to that of a classical reservoir. We use $ U=5\gamma_Q$, random $J$ distributed in $[\pm \gamma_Q/8]$ and a training dataset of $200$ time steps. High decay rate $\gamma_Q\gg J_{mn}$ suppresses the quantum entanglement in the reservoir and thus the quantum advantage is lost.}
\label{LoseQuantum}
\end{figure}
To make use of the larger Hilbert space of the quantum system and access its potentially non-classical correlations, we consider the quantum entanglement measures $S_{mn}$ for the continuous variables between two sites $m$ and $n$\cite{Duan2000,Simon2000} as additional measureable quantities,
\begin{equation}
S_{mn}=V(\hat{p}_m-\hat{p}_n) + V(\hat{q}_m+\hat{q}_n)
\end{equation}
where the amplitude operator $\hat{p}_n =(\hat{a}_n+\hat{a}_n^\dagger)/2 $ and the phase operator $\hat{q}_n =(\hat{a}_n - \hat{a}_n^\dagger)/(2i) $ and the variance of an operator $V(\hat{\mathcal{O}}) = \langle\hat{\mathcal{O}}^2\rangle -\langle\hat{\mathcal{O}}\rangle^2$. Assuming that these quantities are experimentally accessible, we define,
\begin{eqnarray}
y^\text{out}(t) =  \sum_n W^\text{out}_n \langle \hat{a}_n^\dagger(t) \hat{a}_n(t)\rangle \notag\\ + \sum_{mn} D_{nm}^\text{out} \,S_{mn}(t)
\end{eqnarray}
with the feedback $u= y^\text{out}(t-\Delta t)$.
We can add the same entanglement measures to the output of the classical reservoir. However, $S_{mn}$ remains $1$ for any two sites $n$ and $m$ due to the absence of entanglement in a classical system. We study these systems considering a $2\times 3$ lattice. Given that the Hilbert space of $H$ is large, only this small lattice can be simulated within our available computational resources. 

A small classical system ($2\times 3$) has limited accuracy with a NRSE $\sigma\sim0.5$ (see Fig.~\ref{fig:ClassicalPrediction}). The same real sized quantum system obtains better results ($\sigma\sim0.01$) as shown in Fig.~\ref{fig:QuantumPrediction} due to the larger size of its Hilbert space and larger number of available output quantities. It should be noted that here we consider a system in the strongly interacting regime, to obtain non-classical correlations. We have also neglected feedback caused by the process of measuring the quantum system, which implies the use of many copies of the system as considered in Ref.~\cite{Fujii_HarnessingQuantumDynamicsML}.

As a consistency check, let us now take our quantum reservoir in the regime $\text{max}[ J_{nm}] \ll \gamma_Q$, where the hopping between the sites is weaker than the decay rate $\gamma_Q$. In this regime the quantum entanglement is suppressed and thus we lose the advantage of using the quantum reservoir, see Fig.~\ref{LoseQuantum}.
\newpage
\bibliography{bibliography}

\begin{thebibliography}{58}%
\makeatletter
\providecommand \@ifxundefined [1]{%
 \@ifx{#1\undefined}
}%
\providecommand \@ifnum [1]{%
 \ifnum #1\expandafter \@firstoftwo
 \else \expandafter \@secondoftwo
 \fi
}%
\providecommand \@ifx [1]{%
 \ifx #1\expandafter \@firstoftwo
 \else \expandafter \@secondoftwo
 \fi
}%
\providecommand \natexlab [1]{#1}%
\providecommand \enquote  [1]{``#1''}%
\providecommand \bibnamefont  [1]{#1}%
\providecommand \bibfnamefont [1]{#1}%
\providecommand \citenamefont [1]{#1}%
\providecommand \href@noop [0]{\@secondoftwo}%
\providecommand \href [0]{\begingroup \@sanitize@url \@href}%
\providecommand \@href[1]{\@@startlink{#1}\@@href}%
\providecommand \@@href[1]{\endgroup#1\@@endlink}%
\providecommand \@sanitize@url [0]{\catcode `\\12\catcode `\$12\catcode
  `\&12\catcode `\#12\catcode `\^12\catcode `\_12\catcode `\%12\relax}%
\providecommand \@@startlink[1]{}%
\providecommand \@@endlink[0]{}%
\providecommand \url  [0]{\begingroup\@sanitize@url \@url }%
\providecommand \@url [1]{\endgroup\@href {#1}{\urlprefix }}%
\providecommand \urlprefix  [0]{URL }%
\providecommand \Eprint [0]{\href }%
\providecommand \doibase [0]{http://dx.doi.org/}%
\providecommand \selectlanguage [0]{\@gobble}%
\providecommand \bibinfo  [0]{\@secondoftwo}%
\providecommand \bibfield  [0]{\@secondoftwo}%
\providecommand \translation [1]{[#1]}%
\providecommand \BibitemOpen [0]{}%
\providecommand \bibitemStop [0]{}%
\providecommand \bibitemNoStop [0]{.\EOS\space}%
\providecommand \EOS [0]{\spacefactor3000\relax}%
\providecommand \BibitemShut  [1]{\csname bibitem#1\endcsname}%
\let\auto@bib@innerbib\@empty
\bibitem [{\citenamefont {{LeCun}}\ \emph {et~al.}(2015)\citenamefont
  {{LeCun}}, \citenamefont {{Bengio}},\ and\ \citenamefont
  {{Hinton}}}]{LeCun_DeepLearning}%
  \BibitemOpen
  \bibfield  {author} {\bibinfo {author} {\bibfnamefont {Y.}~\bibnamefont
  {{LeCun}}}, \bibinfo {author} {\bibfnamefont {Y.}~\bibnamefont {{Bengio}}}, \
  and\ \bibinfo {author} {\bibfnamefont {G.}~\bibnamefont {{Hinton}}},\ }\href
  {\doibase 10.1038/nature14539} {\bibfield  {journal} {\bibinfo  {journal}
  {\nat}\ }\textbf {\bibinfo {volume} {521}},\ \bibinfo {pages} {436} (\bibinfo
  {year} {2015})}\BibitemShut {NoStop}%
\bibitem [{\citenamefont {{Jaeger}}\ and\ \citenamefont
  {{Haas}}(2004)}]{Jaeger_HarnessingESN}%
  \BibitemOpen
  \bibfield  {author} {\bibinfo {author} {\bibfnamefont {H.}~\bibnamefont
  {{Jaeger}}}\ and\ \bibinfo {author} {\bibfnamefont {H.}~\bibnamefont
  {{Haas}}},\ }\href {\doibase 10.1126/science.1091277} {\bibfield  {journal}
  {\bibinfo  {journal} {Science}\ }\textbf {\bibinfo {volume} {304}},\ \bibinfo
  {pages} {78} (\bibinfo {year} {2004})}\BibitemShut {NoStop}%
\bibitem [{\citenamefont {Maass}\ \emph {et~al.}(2002)\citenamefont {Maass},
  \citenamefont {Natschlager},\ and\ \citenamefont
  {Markram}}]{Maass_RealTimeComputing}%
  \BibitemOpen
  \bibfield  {author} {\bibinfo {author} {\bibfnamefont {W.}~\bibnamefont
  {Maass}}, \bibinfo {author} {\bibfnamefont {T.}~\bibnamefont {Natschlager}},
  \ and\ \bibinfo {author} {\bibfnamefont {H.}~\bibnamefont {Markram}},\ }\href
  {\doibase 10.1162/089976602760407955} {\bibfield  {journal} {\bibinfo
  {journal} {Neural Computation}\ }\textbf {\bibinfo {volume} {14}},\ \bibinfo
  {pages} {2531} (\bibinfo {year} {2002})},\ \Eprint
  {http://arxiv.org/abs/https://doi.org/10.1162/089976602760407955}
  {https://doi.org/10.1162/089976602760407955} \BibitemShut {NoStop}%
\bibitem [{\citenamefont {{Enel}}\ \emph {et~al.}(2016)\citenamefont {{Enel}},
  \citenamefont {{Procyk}}, \citenamefont {{Quilodran}},\ and\ \citenamefont
  {{Dominey}}}]{Enel_RCPropertiesPrefrontalCortex}%
  \BibitemOpen
  \bibfield  {author} {\bibinfo {author} {\bibfnamefont {P.}~\bibnamefont
  {{Enel}}}, \bibinfo {author} {\bibfnamefont {E.}~\bibnamefont {{Procyk}}},
  \bibinfo {author} {\bibfnamefont {R.}~\bibnamefont {{Quilodran}}}, \ and\
  \bibinfo {author} {\bibfnamefont {P.~F.}\ \bibnamefont {{Dominey}}},\ }\href
  {\doibase 10.1371/journal.pcbi.1004967} {\bibfield  {journal} {\bibinfo
  {journal} {PLoS Computational Biology}\ }\textbf {\bibinfo {volume} {12}},\
  \bibinfo {pages} {e1004967} (\bibinfo {year} {2016})}\BibitemShut {NoStop}%
\bibitem [{\citenamefont {Luko{\v{s}}evi{\v{c}}ius}\ and\ \citenamefont
  {Jaeger}(2009)}]{Lukosevicius_RCapproachestoRNN}%
  \BibitemOpen
  \bibfield  {author} {\bibinfo {author} {\bibfnamefont {M.}~\bibnamefont
  {Luko{\v{s}}evi{\v{c}}ius}}\ and\ \bibinfo {author} {\bibfnamefont
  {H.}~\bibnamefont {Jaeger}},\ }\href {\doibase
  https://doi.org/10.1016/j.cosrev.2009.03.005} {\bibfield  {journal} {\bibinfo
   {journal} {Computer Science Review}\ }\textbf {\bibinfo {volume} {3}},\
  \bibinfo {pages} {127 } (\bibinfo {year} {2009})}\BibitemShut {NoStop}%
\bibitem [{\citenamefont
  {Luko{\v{s}}evi{\v{c}}ius}(2012)}]{Lukosevicius_PracticalGuideESN}%
  \BibitemOpen
  \bibfield  {author} {\bibinfo {author} {\bibfnamefont {M.}~\bibnamefont
  {Luko{\v{s}}evi{\v{c}}ius}},\ }\enquote {\bibinfo {title} {A practical guide
  to applying echo state networks},}\ in\ \href {\doibase
  10.1007/978-3-642-35289-8_36} {\emph {\bibinfo {booktitle} {Neural Networks:
  Tricks of the Trade: Second Edition}}},\ \bibinfo {editor} {edited by\
  \bibinfo {editor} {\bibfnamefont {G.}~\bibnamefont {Montavon}}, \bibinfo
  {editor} {\bibfnamefont {G.~B.}\ \bibnamefont {Orr}}, \ and\ \bibinfo
  {editor} {\bibfnamefont {K.-R.}\ \bibnamefont {M{\"u}ller}}}\ (\bibinfo
  {publisher} {Springer Berlin Heidelberg},\ \bibinfo {address} {Berlin,
  Heidelberg},\ \bibinfo {year} {2012})\ pp.\ \bibinfo {pages}
  {659--686}\BibitemShut {NoStop}%
\bibitem [{\citenamefont {{Larger}}\ \emph {et~al.}(2017)\citenamefont
  {{Larger}}, \citenamefont {{Bayl{\'o}n-Fuentes}}, \citenamefont
  {{Martinenghi}}, \citenamefont {{Udaltsov}}, \citenamefont {{Chembo}},\ and\
  \citenamefont {{Jacquot}}}]{Larger_HiSpeedReservoirComputing}%
  \BibitemOpen
  \bibfield  {author} {\bibinfo {author} {\bibfnamefont {L.}~\bibnamefont
  {{Larger}}}, \bibinfo {author} {\bibfnamefont {A.}~\bibnamefont
  {{Bayl{\'o}n-Fuentes}}}, \bibinfo {author} {\bibfnamefont {R.}~\bibnamefont
  {{Martinenghi}}}, \bibinfo {author} {\bibfnamefont {V.~S.}\ \bibnamefont
  {{Udaltsov}}}, \bibinfo {author} {\bibfnamefont {Y.~K.}\ \bibnamefont
  {{Chembo}}}, \ and\ \bibinfo {author} {\bibfnamefont {M.}~\bibnamefont
  {{Jacquot}}},\ }\href {\doibase 10.1103/PhysRevX.7.011015} {\bibfield
  {journal} {\bibinfo  {journal} {Physical Review X}\ }\textbf {\bibinfo
  {volume} {7}},\ \bibinfo {eid} {011015} (\bibinfo {year} {2017})}\BibitemShut
  {NoStop}%
\bibitem [{\citenamefont {{Brunner}}\ \emph {et~al.}(2013)\citenamefont
  {{Brunner}}, \citenamefont {{Soriano}}, \citenamefont {{Mirasso}},\ and\
  \citenamefont {{Fischer}}}]{Brunner_ParallelIPGigabyte}%
  \BibitemOpen
  \bibfield  {author} {\bibinfo {author} {\bibfnamefont {D.}~\bibnamefont
  {{Brunner}}}, \bibinfo {author} {\bibfnamefont {M.~C.}\ \bibnamefont
  {{Soriano}}}, \bibinfo {author} {\bibfnamefont {C.~R.}\ \bibnamefont
  {{Mirasso}}}, \ and\ \bibinfo {author} {\bibfnamefont {I.}~\bibnamefont
  {{Fischer}}},\ }\href {\doibase 10.1038/ncomms2368} {\bibfield  {journal}
  {\bibinfo  {journal} {Nature Communications}\ }\textbf {\bibinfo {volume}
  {4}},\ \bibinfo {eid} {1364} (\bibinfo {year} {2013})}\BibitemShut {NoStop}%
\bibitem [{\citenamefont {{Appeltant}}\ \emph {et~al.}(2011)\citenamefont
  {{Appeltant}}, \citenamefont {{Soriano}}, \citenamefont {{van der Sande}},
  \citenamefont {{Danckaert}}, \citenamefont {{Massar}}, \citenamefont
  {{Dambre}}, \citenamefont {{Schrauwen}}, \citenamefont {{Mirasso}},\ and\
  \citenamefont {{Fischer}}}]{Fischer_InformationProcessingSingleNode}%
  \BibitemOpen
  \bibfield  {author} {\bibinfo {author} {\bibfnamefont {L.}~\bibnamefont
  {{Appeltant}}}, \bibinfo {author} {\bibfnamefont {M.~C.}\ \bibnamefont
  {{Soriano}}}, \bibinfo {author} {\bibfnamefont {G.}~\bibnamefont {{van der
  Sande}}}, \bibinfo {author} {\bibfnamefont {J.}~\bibnamefont {{Danckaert}}},
  \bibinfo {author} {\bibfnamefont {S.}~\bibnamefont {{Massar}}}, \bibinfo
  {author} {\bibfnamefont {J.}~\bibnamefont {{Dambre}}}, \bibinfo {author}
  {\bibfnamefont {B.}~\bibnamefont {{Schrauwen}}}, \bibinfo {author}
  {\bibfnamefont {C.~R.}\ \bibnamefont {{Mirasso}}}, \ and\ \bibinfo {author}
  {\bibfnamefont {I.}~\bibnamefont {{Fischer}}},\ }\href {\doibase
  10.1038/ncomms1476} {\bibfield  {journal} {\bibinfo  {journal} {Nature
  Communications}\ }\textbf {\bibinfo {volume} {2}},\ \bibinfo {eid} {468}
  (\bibinfo {year} {2011})}\BibitemShut {NoStop}%
\bibitem [{\citenamefont {{Fujii}}\ and\ \citenamefont
  {{Nakajima}}(2017)}]{Fujii_HarnessingQuantumDynamicsML}%
  \BibitemOpen
  \bibfield  {author} {\bibinfo {author} {\bibfnamefont {K.}~\bibnamefont
  {{Fujii}}}\ and\ \bibinfo {author} {\bibfnamefont {K.}~\bibnamefont
  {{Nakajima}}},\ }\href {\doibase 10.1103/PhysRevApplied.8.024030} {\bibfield
  {journal} {\bibinfo  {journal} {Physical Review Applied}\ }\textbf {\bibinfo
  {volume} {8}},\ \bibinfo {eid} {024030} (\bibinfo {year} {2017})},\ \Eprint
  {http://arxiv.org/abs/1602.08159} {arXiv:1602.08159 [quant-ph]} \BibitemShut
  {NoStop}%
\bibitem [{\citenamefont {{Vandoorne}}\ \emph {et~al.}(2014)\citenamefont
  {{Vandoorne}}, \citenamefont {{Mechet}}, \citenamefont {{van Vaerenbergh}},
  \citenamefont {{Fiers}}, \citenamefont {{Morthier}}, \citenamefont
  {{Verstraeten}}, \citenamefont {{Schrauwen}}, \citenamefont {{Dambre}},\ and\
  \citenamefont {{Bienstman}}}]{Vandoorne_RCSiliconChip}%
  \BibitemOpen
  \bibfield  {author} {\bibinfo {author} {\bibfnamefont {K.}~\bibnamefont
  {{Vandoorne}}}, \bibinfo {author} {\bibfnamefont {P.}~\bibnamefont
  {{Mechet}}}, \bibinfo {author} {\bibfnamefont {T.}~\bibnamefont {{van
  Vaerenbergh}}}, \bibinfo {author} {\bibfnamefont {M.}~\bibnamefont
  {{Fiers}}}, \bibinfo {author} {\bibfnamefont {G.}~\bibnamefont {{Morthier}}},
  \bibinfo {author} {\bibfnamefont {D.}~\bibnamefont {{Verstraeten}}}, \bibinfo
  {author} {\bibfnamefont {B.}~\bibnamefont {{Schrauwen}}}, \bibinfo {author}
  {\bibfnamefont {J.}~\bibnamefont {{Dambre}}}, \ and\ \bibinfo {author}
  {\bibfnamefont {P.}~\bibnamefont {{Bienstman}}},\ }\href {\doibase
  10.1038/ncomms4541} {\bibfield  {journal} {\bibinfo  {journal} {Nature
  Communications}\ }\textbf {\bibinfo {volume} {5}},\ \bibinfo {eid} {3541}
  (\bibinfo {year} {2014})}\BibitemShut {NoStop}%
\bibitem [{\citenamefont {Torrejon}\ \emph {et~al.}(2017)\citenamefont
  {Torrejon}, \citenamefont {Riou}, \citenamefont {Araujo}, \citenamefont
  {Tsunegi}, \citenamefont {Khalsa}, \citenamefont {Querlioz}, \citenamefont
  {Bortolotti}, \citenamefont {Cros}, \citenamefont {Yakushiji}, \citenamefont
  {Fukushima}, \citenamefont {Kubota}, \citenamefont {Yuasa}, \citenamefont
  {Stiles},\ and\ \citenamefont {Grollier}}]{Torrejon_Neuromorphic}%
  \BibitemOpen
  \bibfield  {author} {\bibinfo {author} {\bibfnamefont {J.}~\bibnamefont
  {Torrejon}}, \bibinfo {author} {\bibfnamefont {M.}~\bibnamefont {Riou}},
  \bibinfo {author} {\bibfnamefont {F.~A.}\ \bibnamefont {Araujo}}, \bibinfo
  {author} {\bibfnamefont {S.}~\bibnamefont {Tsunegi}}, \bibinfo {author}
  {\bibfnamefont {G.}~\bibnamefont {Khalsa}}, \bibinfo {author} {\bibfnamefont
  {D.}~\bibnamefont {Querlioz}}, \bibinfo {author} {\bibfnamefont
  {P.}~\bibnamefont {Bortolotti}}, \bibinfo {author} {\bibfnamefont
  {V.}~\bibnamefont {Cros}}, \bibinfo {author} {\bibfnamefont {K.}~\bibnamefont
  {Yakushiji}}, \bibinfo {author} {\bibfnamefont {A.}~\bibnamefont
  {Fukushima}}, \bibinfo {author} {\bibfnamefont {H.}~\bibnamefont {Kubota}},
  \bibinfo {author} {\bibfnamefont {S.}~\bibnamefont {Yuasa}}, \bibinfo
  {author} {\bibfnamefont {M.~D.}\ \bibnamefont {Stiles}}, \ and\ \bibinfo
  {author} {\bibfnamefont {J.}~\bibnamefont {Grollier}},\ }\href@noop {}
  {\bibfield  {journal} {\bibinfo  {journal} {Nature}\ }\textbf {\bibinfo
  {volume} {547}},\ \bibinfo {pages} {428} (\bibinfo {year}
  {2017})}\BibitemShut {NoStop}%
\bibitem [{Jae()}]{Jaeger_competition}%
  \BibitemOpen
  \href@noop {} {}\bibinfo {howpublished}
  {\url{http://www.neural-forecasting-competition.com/NN3/index.htm}}\BibitemShut
  {NoStop}%
\bibitem [{\citenamefont {Sch\"{u}rmann}\ \emph {et~al.}(2005)\citenamefont
  {Sch\"{u}rmann}, \citenamefont {Meier},\ and\ \citenamefont
  {Schemmel}}]{VLSI_Chip}%
  \BibitemOpen
  \bibfield  {author} {\bibinfo {author} {\bibfnamefont {F.}~\bibnamefont
  {Sch\"{u}rmann}}, \bibinfo {author} {\bibfnamefont {K.}~\bibnamefont
  {Meier}}, \ and\ \bibinfo {author} {\bibfnamefont {J.}~\bibnamefont
  {Schemmel}},\ }in\ \href
  {http://papers.nips.cc/paper/2562-edge-of-chaos-computation-in-mixed-mode-vlsi-a-hard-liquid.pdf}
  {\emph {\bibinfo {booktitle} {Advances in Neural Information Processing
  Systems 17}}},\ \bibinfo {editor} {edited by\ \bibinfo {editor}
  {\bibfnamefont {L.~K.}\ \bibnamefont {Saul}}, \bibinfo {editor}
  {\bibfnamefont {Y.}~\bibnamefont {Weiss}}, \ and\ \bibinfo {editor}
  {\bibfnamefont {L.}~\bibnamefont {Bottou}}}\ (\bibinfo  {publisher} {MIT
  Press},\ \bibinfo {year} {2005})\ pp.\ \bibinfo {pages}
  {1201--1208}\BibitemShut {NoStop}%
\bibitem [{\citenamefont {{Du}}\ \emph {et~al.}(2017)\citenamefont {{Du}},
  \citenamefont {{Cai}}, \citenamefont {{Zidan}}, \citenamefont {{Ma}},
  \citenamefont {{Lee}},\ and\ \citenamefont {{Lu}}}]{Du_MemristorRC}%
  \BibitemOpen
  \bibfield  {author} {\bibinfo {author} {\bibfnamefont {C.}~\bibnamefont
  {{Du}}}, \bibinfo {author} {\bibfnamefont {F.}~\bibnamefont {{Cai}}},
  \bibinfo {author} {\bibfnamefont {M.~A.}\ \bibnamefont {{Zidan}}}, \bibinfo
  {author} {\bibfnamefont {W.}~\bibnamefont {{Ma}}}, \bibinfo {author}
  {\bibfnamefont {S.~H.}\ \bibnamefont {{Lee}}}, \ and\ \bibinfo {author}
  {\bibfnamefont {W.~D.}\ \bibnamefont {{Lu}}},\ }\href {\doibase
  10.1038/s41467-017-02337-y} {\bibfield  {journal} {\bibinfo  {journal}
  {Nature Communications}\ }\textbf {\bibinfo {volume} {8}},\ \bibinfo {eid}
  {2204} (\bibinfo {year} {2017})}\BibitemShut {NoStop}%
\bibitem [{\citenamefont {Kudithipudi}\ \emph {et~al.}(2016)\citenamefont
  {Kudithipudi}, \citenamefont {Saleh}, \citenamefont {Merkel}, \citenamefont
  {Thesing},\ and\ \citenamefont
  {Wysocki}}]{Kudithipudi_NeuromemristiveRCBiosignalProcessing}%
  \BibitemOpen
  \bibfield  {author} {\bibinfo {author} {\bibfnamefont {D.}~\bibnamefont
  {Kudithipudi}}, \bibinfo {author} {\bibfnamefont {Q.}~\bibnamefont {Saleh}},
  \bibinfo {author} {\bibfnamefont {C.}~\bibnamefont {Merkel}}, \bibinfo
  {author} {\bibfnamefont {J.}~\bibnamefont {Thesing}}, \ and\ \bibinfo
  {author} {\bibfnamefont {B.}~\bibnamefont {Wysocki}},\ }\href {\doibase
  10.3389/fnins.2015.00502} {\bibfield  {journal} {\bibinfo  {journal}
  {Frontiers in Neuroscience}\ }\textbf {\bibinfo {volume} {9}},\ \bibinfo
  {pages} {502} (\bibinfo {year} {2016})}\BibitemShut {NoStop}%
\bibitem [{\citenamefont {{Paquot}}\ \emph {et~al.}(2012)\citenamefont
  {{Paquot}}, \citenamefont {{Duport}}, \citenamefont {{Smerieri}},
  \citenamefont {{Dambre}}, \citenamefont {{Schrauwen}}, \citenamefont
  {{Haelterman}},\ and\ \citenamefont {{Massar}}}]{Paquot_OptoelectronicRC}%
  \BibitemOpen
  \bibfield  {author} {\bibinfo {author} {\bibfnamefont {Y.}~\bibnamefont
  {{Paquot}}}, \bibinfo {author} {\bibfnamefont {F.}~\bibnamefont {{Duport}}},
  \bibinfo {author} {\bibfnamefont {A.}~\bibnamefont {{Smerieri}}}, \bibinfo
  {author} {\bibfnamefont {J.}~\bibnamefont {{Dambre}}}, \bibinfo {author}
  {\bibfnamefont {B.}~\bibnamefont {{Schrauwen}}}, \bibinfo {author}
  {\bibfnamefont {M.}~\bibnamefont {{Haelterman}}}, \ and\ \bibinfo {author}
  {\bibfnamefont {S.}~\bibnamefont {{Massar}}},\ }\href {\doibase
  10.1038/srep00287} {\bibfield  {journal} {\bibinfo  {journal} {Scientific
  Reports}\ }\textbf {\bibinfo {volume} {2}},\ \bibinfo {eid} {287} (\bibinfo
  {year} {2012})},\ \Eprint {http://arxiv.org/abs/1111.7219} {arXiv:1111.7219}
  \BibitemShut {NoStop}%
\bibitem [{\citenamefont {Larger}\ \emph {et~al.}(2012)\citenamefont {Larger},
  \citenamefont {Soriano}, \citenamefont {Brunner}, \citenamefont {Appeltant},
  \citenamefont {Gutierrez}, \citenamefont {Pesquera}, \citenamefont
  {Mirasso},\ and\ \citenamefont {Fischer}}]{Larger_BeyondTuring}%
  \BibitemOpen
  \bibfield  {author} {\bibinfo {author} {\bibfnamefont {L.}~\bibnamefont
  {Larger}}, \bibinfo {author} {\bibfnamefont {M.~C.}\ \bibnamefont {Soriano}},
  \bibinfo {author} {\bibfnamefont {D.}~\bibnamefont {Brunner}}, \bibinfo
  {author} {\bibfnamefont {L.}~\bibnamefont {Appeltant}}, \bibinfo {author}
  {\bibfnamefont {J.~M.}\ \bibnamefont {Gutierrez}}, \bibinfo {author}
  {\bibfnamefont {L.}~\bibnamefont {Pesquera}}, \bibinfo {author}
  {\bibfnamefont {C.~R.}\ \bibnamefont {Mirasso}}, \ and\ \bibinfo {author}
  {\bibfnamefont {I.}~\bibnamefont {Fischer}},\ }\href {\doibase
  10.1364/OE.20.003241} {\bibfield  {journal} {\bibinfo  {journal} {Opt.
  Express}\ }\textbf {\bibinfo {volume} {20}},\ \bibinfo {pages} {3241}
  (\bibinfo {year} {2012})}\BibitemShut {NoStop}%
\bibitem [{\citenamefont {{Duport}}\ \emph {et~al.}(2012)\citenamefont
  {{Duport}}, \citenamefont {{Schneider}}, \citenamefont {{Smerieri}},
  \citenamefont {{Haelterman}},\ and\ \citenamefont
  {{Massar}}}]{Duport_AllOpticalRC}%
  \BibitemOpen
  \bibfield  {author} {\bibinfo {author} {\bibfnamefont {F.}~\bibnamefont
  {{Duport}}}, \bibinfo {author} {\bibfnamefont {B.}~\bibnamefont
  {{Schneider}}}, \bibinfo {author} {\bibfnamefont {A.}~\bibnamefont
  {{Smerieri}}}, \bibinfo {author} {\bibfnamefont {M.}~\bibnamefont
  {{Haelterman}}}, \ and\ \bibinfo {author} {\bibfnamefont {S.}~\bibnamefont
  {{Massar}}},\ }\href {\doibase 10.1364/OE.20.022783} {\bibfield  {journal}
  {\bibinfo  {journal} {Optics Express}\ }\textbf {\bibinfo {volume} {20}},\
  \bibinfo {pages} {22783} (\bibinfo {year} {2012})},\ \Eprint
  {http://arxiv.org/abs/1207.1619} {arXiv:1207.1619 [physics.optics]}
  \BibitemShut {NoStop}%
\bibitem [{\citenamefont {{Coulombe}}\ \emph {et~al.}(2017)\citenamefont
  {{Coulombe}}, \citenamefont {{York}},\ and\ \citenamefont
  {{Sylvestre}}}]{Coulombe_ComputingMechanicalOscillators}%
  \BibitemOpen
  \bibfield  {author} {\bibinfo {author} {\bibfnamefont {J.~C.}\ \bibnamefont
  {{Coulombe}}}, \bibinfo {author} {\bibfnamefont {M.~C.~A.}\ \bibnamefont
  {{York}}}, \ and\ \bibinfo {author} {\bibfnamefont {J.}~\bibnamefont
  {{Sylvestre}}},\ }\href {\doibase 10.1371/journal.pone.0178663} {\bibfield
  {journal} {\bibinfo  {journal} {PLoS ONE}\ }\textbf {\bibinfo {volume}
  {12}},\ \bibinfo {pages} {e0178663} (\bibinfo {year} {2017})},\ \Eprint
  {http://arxiv.org/abs/1704.06320} {arXiv:1704.06320} \BibitemShut {NoStop}%
\bibitem [{\citenamefont {Fernando}\ and\ \citenamefont
  {Sojakka}(2003)}]{Bucket}%
  \BibitemOpen
  \bibfield  {author} {\bibinfo {author} {\bibfnamefont {C.}~\bibnamefont
  {Fernando}}\ and\ \bibinfo {author} {\bibfnamefont {S.}~\bibnamefont
  {Sojakka}},\ }in\ \href@noop {} {\emph {\bibinfo {booktitle} {Advances in
  Artificial Life}}},\ \bibinfo {editor} {edited by\ \bibinfo {editor}
  {\bibfnamefont {W.}~\bibnamefont {Banzhaf}}, \bibinfo {editor} {\bibfnamefont
  {J.}~\bibnamefont {Ziegler}}, \bibinfo {editor} {\bibfnamefont
  {T.}~\bibnamefont {Christaller}}, \bibinfo {editor} {\bibfnamefont
  {P.}~\bibnamefont {Dittrich}}, \ and\ \bibinfo {editor} {\bibfnamefont
  {J.~T.}\ \bibnamefont {Kim}}}\ (\bibinfo  {publisher} {Springer Berlin
  Heidelberg},\ \bibinfo {address} {Berlin, Heidelberg},\ \bibinfo {year}
  {2003})\ pp.\ \bibinfo {pages} {588--597}\BibitemShut {NoStop}%
\bibitem [{\citenamefont {{Tanaka}}\ \emph {et~al.}(2018)\citenamefont
  {{Tanaka}}, \citenamefont {{Yamane}}, \citenamefont {{Benoit H{\'e}roux}},
  \citenamefont {{Nakane}}, \citenamefont {{Kanazawa}}, \citenamefont
  {{Takeda}}, \citenamefont {{Numata}}, \citenamefont {{Nakano}},\ and\
  \citenamefont {{Hirose}}}]{Tanaka_review}%
  \BibitemOpen
  \bibfield  {author} {\bibinfo {author} {\bibfnamefont {G.}~\bibnamefont
  {{Tanaka}}}, \bibinfo {author} {\bibfnamefont {T.}~\bibnamefont {{Yamane}}},
  \bibinfo {author} {\bibfnamefont {J.}~\bibnamefont {{Benoit H{\'e}roux}}},
  \bibinfo {author} {\bibfnamefont {R.}~\bibnamefont {{Nakane}}}, \bibinfo
  {author} {\bibfnamefont {N.}~\bibnamefont {{Kanazawa}}}, \bibinfo {author}
  {\bibfnamefont {S.}~\bibnamefont {{Takeda}}}, \bibinfo {author}
  {\bibfnamefont {H.}~\bibnamefont {{Numata}}}, \bibinfo {author}
  {\bibfnamefont {D.}~\bibnamefont {{Nakano}}}, \ and\ \bibinfo {author}
  {\bibfnamefont {A.}~\bibnamefont {{Hirose}}},\ }\href@noop {} {\bibfield
  {journal} {\bibinfo  {journal} {ArXiv e-prints}\ ,\ \bibinfo {eid}
  {arXiv:1808.04962}} (\bibinfo {year} {2018})},\ \Eprint
  {http://arxiv.org/abs/1808.04962} {arXiv:1808.04962 [cs.ET]} \BibitemShut
  {NoStop}%
\bibitem [{\citenamefont {{Shen}}\ \emph {et~al.}(2017)\citenamefont {{Shen}},
  \citenamefont {{Harris}}, \citenamefont {{Skirlo}}, \citenamefont {{Prabhu}},
  \citenamefont {{Baehr-Jones}}, \citenamefont {{Hochberg}}, \citenamefont
  {{Sun}}, \citenamefont {{Zhao}}, \citenamefont {{Larochelle}}, \citenamefont
  {{Englund}},\ and\ \citenamefont {{Solja{\v
  c}i{\'c}}}}]{Soljacic_DeepLearning}%
  \BibitemOpen
  \bibfield  {author} {\bibinfo {author} {\bibfnamefont {Y.}~\bibnamefont
  {{Shen}}}, \bibinfo {author} {\bibfnamefont {N.~C.}\ \bibnamefont
  {{Harris}}}, \bibinfo {author} {\bibfnamefont {S.}~\bibnamefont {{Skirlo}}},
  \bibinfo {author} {\bibfnamefont {M.}~\bibnamefont {{Prabhu}}}, \bibinfo
  {author} {\bibfnamefont {T.}~\bibnamefont {{Baehr-Jones}}}, \bibinfo {author}
  {\bibfnamefont {M.}~\bibnamefont {{Hochberg}}}, \bibinfo {author}
  {\bibfnamefont {X.}~\bibnamefont {{Sun}}}, \bibinfo {author} {\bibfnamefont
  {S.}~\bibnamefont {{Zhao}}}, \bibinfo {author} {\bibfnamefont
  {H.}~\bibnamefont {{Larochelle}}}, \bibinfo {author} {\bibfnamefont
  {D.}~\bibnamefont {{Englund}}}, \ and\ \bibinfo {author} {\bibfnamefont
  {M.}~\bibnamefont {{Solja{\v c}i{\'c}}}},\ }\href {\doibase
  10.1038/nphoton.2017.93} {\bibfield  {journal} {\bibinfo  {journal} {Nature
  Photonics}\ }\textbf {\bibinfo {volume} {11}},\ \bibinfo {pages} {441}
  (\bibinfo {year} {2017})},\ \Eprint {http://arxiv.org/abs/1610.02365}
  {arXiv:1610.02365 [physics.optics]} \BibitemShut {NoStop}%
\bibitem [{\citenamefont {Caulfield}\ and\ \citenamefont
  {Dolev}(2010)}]{Caulfield_FutureSupercomputing}%
  \BibitemOpen
  \bibfield  {author} {\bibinfo {author} {\bibfnamefont {H.~J.}\ \bibnamefont
  {Caulfield}}\ and\ \bibinfo {author} {\bibfnamefont {S.}~\bibnamefont
  {Dolev}},\ }\href {http://dx.doi.org/10.1038/nphoton.2010.94} {\bibfield
  {journal} {\bibinfo  {journal} {Nature Photonics}\ }\textbf {\bibinfo
  {volume} {4}},\ \bibinfo {pages} {261 EP } (\bibinfo {year}
  {2010})}\BibitemShut {NoStop}%
\bibitem [{\citenamefont {Aranson}\ and\ \citenamefont
  {Kramer}(2002)}]{Aranson_CGLEWorld}%
  \BibitemOpen
  \bibfield  {author} {\bibinfo {author} {\bibfnamefont {I.~S.}\ \bibnamefont
  {Aranson}}\ and\ \bibinfo {author} {\bibfnamefont {L.}~\bibnamefont
  {Kramer}},\ }\href {\doibase 10.1103/RevModPhys.74.99} {\bibfield  {journal}
  {\bibinfo  {journal} {Rev. Mod. Phys.}\ }\textbf {\bibinfo {volume} {74}},\
  \bibinfo {pages} {99} (\bibinfo {year} {2002})}\BibitemShut {NoStop}%
\bibitem [{\citenamefont {Keeling}\ and\ \citenamefont
  {Berloff}(2008)}]{Berloff_VortexLattices}%
  \BibitemOpen
  \bibfield  {author} {\bibinfo {author} {\bibfnamefont {J.}~\bibnamefont
  {Keeling}}\ and\ \bibinfo {author} {\bibfnamefont {N.~G.}\ \bibnamefont
  {Berloff}},\ }\href {\doibase 10.1103/PhysRevLett.100.250401} {\bibfield
  {journal} {\bibinfo  {journal} {Phys. Rev. Lett.}\ }\textbf {\bibinfo
  {volume} {100}},\ \bibinfo {pages} {250401} (\bibinfo {year}
  {2008})}\BibitemShut {NoStop}%
\bibitem [{\citenamefont {Labouvie}\ \emph {et~al.}(2016)\citenamefont
  {Labouvie}, \citenamefont {Santra}, \citenamefont {Heun},\ and\ \citenamefont
  {Ott}}]{Ott_BistabilityDrivenDissipative}%
  \BibitemOpen
  \bibfield  {author} {\bibinfo {author} {\bibfnamefont {R.}~\bibnamefont
  {Labouvie}}, \bibinfo {author} {\bibfnamefont {B.}~\bibnamefont {Santra}},
  \bibinfo {author} {\bibfnamefont {S.}~\bibnamefont {Heun}}, \ and\ \bibinfo
  {author} {\bibfnamefont {H.}~\bibnamefont {Ott}},\ }\href {\doibase
  10.1103/PhysRevLett.116.235302} {\bibfield  {journal} {\bibinfo  {journal}
  {Phys. Rev. Lett.}\ }\textbf {\bibinfo {volume} {116}},\ \bibinfo {pages}
  {235302} (\bibinfo {year} {2016})}\BibitemShut {NoStop}%
\bibitem [{\citenamefont {Walker}\ \emph {et~al.}(2015)\citenamefont {Walker},
  \citenamefont {Tinkler}, \citenamefont {Skryabin}, \citenamefont {Yulin},
  \citenamefont {Royall}, \citenamefont {Farrer}, \citenamefont {Ritchie},
  \citenamefont {Skolnick},\ and\ \citenamefont
  {Krizhanovskii}}]{Krizhanovskii_UltralowPowerSolitons}%
  \BibitemOpen
  \bibfield  {author} {\bibinfo {author} {\bibfnamefont {P.~M.}\ \bibnamefont
  {Walker}}, \bibinfo {author} {\bibfnamefont {L.}~\bibnamefont {Tinkler}},
  \bibinfo {author} {\bibfnamefont {D.~V.}\ \bibnamefont {Skryabin}}, \bibinfo
  {author} {\bibfnamefont {A.}~\bibnamefont {Yulin}}, \bibinfo {author}
  {\bibfnamefont {B.}~\bibnamefont {Royall}}, \bibinfo {author} {\bibfnamefont
  {I.}~\bibnamefont {Farrer}}, \bibinfo {author} {\bibfnamefont {D.~A.}\
  \bibnamefont {Ritchie}}, \bibinfo {author} {\bibfnamefont {M.~S.}\
  \bibnamefont {Skolnick}}, \ and\ \bibinfo {author} {\bibfnamefont {D.~N.}\
  \bibnamefont {Krizhanovskii}},\ }\href {http://dx.doi.org/10.1038/ncomms9317}
  {\bibfield  {journal} {\bibinfo  {journal} {Nature Communications}\ }\textbf
  {\bibinfo {volume} {6}},\ \bibinfo {pages} {8317 EP } (\bibinfo {year}
  {2015})},\ \bibinfo {note} {article}\BibitemShut {NoStop}%
\bibitem [{\citenamefont {Kim}\ \emph {et~al.}(2013)\citenamefont {Kim},
  \citenamefont {Kusudo}, \citenamefont {Löffler}, \citenamefont {Höfling},
  \citenamefont {Forchel},\ and\ \citenamefont {Yamamoto}}]{Kim_2013}%
  \BibitemOpen
  \bibfield  {author} {\bibinfo {author} {\bibfnamefont {N.~Y.}\ \bibnamefont
  {Kim}}, \bibinfo {author} {\bibfnamefont {K.}~\bibnamefont {Kusudo}},
  \bibinfo {author} {\bibfnamefont {A.}~\bibnamefont {Löffler}}, \bibinfo
  {author} {\bibfnamefont {S.}~\bibnamefont {Höfling}}, \bibinfo {author}
  {\bibfnamefont {A.}~\bibnamefont {Forchel}}, \ and\ \bibinfo {author}
  {\bibfnamefont {Y.}~\bibnamefont {Yamamoto}},\ }\href {\doibase
  10.1088/1367-2630/15/3/035032} {\bibfield  {journal} {\bibinfo  {journal}
  {New Journal of Physics}\ }\textbf {\bibinfo {volume} {15}},\ \bibinfo
  {pages} {035032} (\bibinfo {year} {2013})}\BibitemShut {NoStop}%
\bibitem [{\citenamefont {Winkler}\ \emph {et~al.}(2016)\citenamefont
  {Winkler}, \citenamefont {Egorov}, \citenamefont {Savenko}, \citenamefont
  {Ma}, \citenamefont {Estrecho}, \citenamefont {Gao}, \citenamefont
  {M\"uller}, \citenamefont {Kamp}, \citenamefont {Liew}, \citenamefont
  {Ostrovskaya}, \citenamefont {H\"ofling},\ and\ \citenamefont
  {Schneider}}]{Winkler_2016}%
  \BibitemOpen
  \bibfield  {author} {\bibinfo {author} {\bibfnamefont {K.}~\bibnamefont
  {Winkler}}, \bibinfo {author} {\bibfnamefont {O.~A.}\ \bibnamefont {Egorov}},
  \bibinfo {author} {\bibfnamefont {I.~G.}\ \bibnamefont {Savenko}}, \bibinfo
  {author} {\bibfnamefont {X.}~\bibnamefont {Ma}}, \bibinfo {author}
  {\bibfnamefont {E.}~\bibnamefont {Estrecho}}, \bibinfo {author}
  {\bibfnamefont {T.}~\bibnamefont {Gao}}, \bibinfo {author} {\bibfnamefont
  {S.}~\bibnamefont {M\"uller}}, \bibinfo {author} {\bibfnamefont
  {M.}~\bibnamefont {Kamp}}, \bibinfo {author} {\bibfnamefont {T.~C.~H.}\
  \bibnamefont {Liew}}, \bibinfo {author} {\bibfnamefont {E.~A.}\ \bibnamefont
  {Ostrovskaya}}, \bibinfo {author} {\bibfnamefont {S.}~\bibnamefont
  {H\"ofling}}, \ and\ \bibinfo {author} {\bibfnamefont {C.}~\bibnamefont
  {Schneider}},\ }\href {\doibase 10.1103/PhysRevB.93.121303} {\bibfield
  {journal} {\bibinfo  {journal} {Phys. Rev. B}\ }\textbf {\bibinfo {volume}
  {93}},\ \bibinfo {pages} {121303} (\bibinfo {year} {2016})}\BibitemShut
  {NoStop}%
\bibitem [{\citenamefont {Sala}\ \emph {et~al.}(2015)\citenamefont {Sala},
  \citenamefont {Solnyshkov}, \citenamefont {Carusotto}, \citenamefont
  {Jacqmin}, \citenamefont {Lema\^{\i}tre}, \citenamefont
  {Ter\ifmmode~\mbox{\c{c}}\else \c{c}\fi{}as}, \citenamefont {Nalitov},
  \citenamefont {Abbarchi}, \citenamefont {Galopin}, \citenamefont {Sagnes},
  \citenamefont {Bloch}, \citenamefont {Malpuech},\ and\ \citenamefont
  {Amo}}]{Amo_SOCoupling}%
  \BibitemOpen
  \bibfield  {author} {\bibinfo {author} {\bibfnamefont {V.~G.}\ \bibnamefont
  {Sala}}, \bibinfo {author} {\bibfnamefont {D.~D.}\ \bibnamefont
  {Solnyshkov}}, \bibinfo {author} {\bibfnamefont {I.}~\bibnamefont
  {Carusotto}}, \bibinfo {author} {\bibfnamefont {T.}~\bibnamefont {Jacqmin}},
  \bibinfo {author} {\bibfnamefont {A.}~\bibnamefont {Lema\^{\i}tre}}, \bibinfo
  {author} {\bibfnamefont {H.}~\bibnamefont {Ter\ifmmode~\mbox{\c{c}}\else
  \c{c}\fi{}as}}, \bibinfo {author} {\bibfnamefont {A.}~\bibnamefont
  {Nalitov}}, \bibinfo {author} {\bibfnamefont {M.}~\bibnamefont {Abbarchi}},
  \bibinfo {author} {\bibfnamefont {E.}~\bibnamefont {Galopin}}, \bibinfo
  {author} {\bibfnamefont {I.}~\bibnamefont {Sagnes}}, \bibinfo {author}
  {\bibfnamefont {J.}~\bibnamefont {Bloch}}, \bibinfo {author} {\bibfnamefont
  {G.}~\bibnamefont {Malpuech}}, \ and\ \bibinfo {author} {\bibfnamefont
  {A.}~\bibnamefont {Amo}},\ }\href {\doibase 10.1103/PhysRevX.5.011034}
  {\bibfield  {journal} {\bibinfo  {journal} {Phys. Rev. X}\ }\textbf {\bibinfo
  {volume} {5}},\ \bibinfo {pages} {011034} (\bibinfo {year}
  {2015})}\BibitemShut {NoStop}%
\bibitem [{\citenamefont {Whittaker}\ \emph {et~al.}(2018)\citenamefont
  {Whittaker}, \citenamefont {Cancellieri}, \citenamefont {Walker},
  \citenamefont {Gulevich}, \citenamefont {Schomerus}, \citenamefont
  {Vaitiekus}, \citenamefont {Royall}, \citenamefont {Whittaker}, \citenamefont
  {Clarke}, \citenamefont {Iorsh}, \citenamefont {Shelykh}, \citenamefont
  {Skolnick},\ and\ \citenamefont {Krizhanovskii}}]{Krizhanovskii_LiebLattice}%
  \BibitemOpen
  \bibfield  {author} {\bibinfo {author} {\bibfnamefont {C.~E.}\ \bibnamefont
  {Whittaker}}, \bibinfo {author} {\bibfnamefont {E.}~\bibnamefont
  {Cancellieri}}, \bibinfo {author} {\bibfnamefont {P.~M.}\ \bibnamefont
  {Walker}}, \bibinfo {author} {\bibfnamefont {D.~R.}\ \bibnamefont
  {Gulevich}}, \bibinfo {author} {\bibfnamefont {H.}~\bibnamefont {Schomerus}},
  \bibinfo {author} {\bibfnamefont {D.}~\bibnamefont {Vaitiekus}}, \bibinfo
  {author} {\bibfnamefont {B.}~\bibnamefont {Royall}}, \bibinfo {author}
  {\bibfnamefont {D.~M.}\ \bibnamefont {Whittaker}}, \bibinfo {author}
  {\bibfnamefont {E.}~\bibnamefont {Clarke}}, \bibinfo {author} {\bibfnamefont
  {I.~V.}\ \bibnamefont {Iorsh}}, \bibinfo {author} {\bibfnamefont {I.~A.}\
  \bibnamefont {Shelykh}}, \bibinfo {author} {\bibfnamefont {M.~S.}\
  \bibnamefont {Skolnick}}, \ and\ \bibinfo {author} {\bibfnamefont {D.~N.}\
  \bibnamefont {Krizhanovskii}},\ }\href {\doibase
  10.1103/PhysRevLett.120.097401} {\bibfield  {journal} {\bibinfo  {journal}
  {Phys. Rev. Lett.}\ }\textbf {\bibinfo {volume} {120}},\ \bibinfo {pages}
  {097401} (\bibinfo {year} {2018})}\BibitemShut {NoStop}%
\bibitem [{\citenamefont {Hopfield}(1958)}]{Hopfield}%
  \BibitemOpen
  \bibfield  {author} {\bibinfo {author} {\bibfnamefont {J.~J.}\ \bibnamefont
  {Hopfield}},\ }\href {\doibase 10.1103/PhysRev.112.1555} {\bibfield
  {journal} {\bibinfo  {journal} {Phys. Rev.}\ }\textbf {\bibinfo {volume}
  {112}},\ \bibinfo {pages} {1555} (\bibinfo {year} {1958})}\BibitemShut
  {NoStop}%
\bibitem [{\citenamefont {Weisbuch}\ \emph {et~al.}(1992)\citenamefont
  {Weisbuch}, \citenamefont {Nishioka}, \citenamefont {Ishikawa},\ and\
  \citenamefont {Arakawa}}]{Weisbuch}%
  \BibitemOpen
  \bibfield  {author} {\bibinfo {author} {\bibfnamefont {C.}~\bibnamefont
  {Weisbuch}}, \bibinfo {author} {\bibfnamefont {M.}~\bibnamefont {Nishioka}},
  \bibinfo {author} {\bibfnamefont {A.}~\bibnamefont {Ishikawa}}, \ and\
  \bibinfo {author} {\bibfnamefont {Y.}~\bibnamefont {Arakawa}},\ }\href
  {\doibase 10.1103/PhysRevLett.69.3314} {\bibfield  {journal} {\bibinfo
  {journal} {Phys. Rev. Lett.}\ }\textbf {\bibinfo {volume} {69}},\ \bibinfo
  {pages} {3314} (\bibinfo {year} {1992})}\BibitemShut {NoStop}%
\bibitem [{\citenamefont {Kavokin}\ \emph {et~al.}(2008)\citenamefont
  {Kavokin}, \citenamefont {Baumberg}, \citenamefont {Malpuech},\ and\
  \citenamefont {Laussy}}]{Kavokin}%
  \BibitemOpen
  \bibfield  {author} {\bibinfo {author} {\bibfnamefont {A.}~\bibnamefont
  {Kavokin}}, \bibinfo {author} {\bibfnamefont {J.~J.}\ \bibnamefont
  {Baumberg}}, \bibinfo {author} {\bibfnamefont {G.}~\bibnamefont {Malpuech}},
  \ and\ \bibinfo {author} {\bibfnamefont {F.~P.}\ \bibnamefont {Laussy}},\
  }\href@noop {} {\emph {\bibinfo {title} {Microcavities}}}\ (\bibinfo
  {publisher} {Oxford University Press, Inc.},\ \bibinfo {address} {New York,
  NY, USA},\ \bibinfo {year} {2008})\BibitemShut {NoStop}%
\bibitem [{\citenamefont {Carusotto}\ and\ \citenamefont
  {Ciuti}(2013)}]{Carusotto_QuantumFluids}%
  \BibitemOpen
  \bibfield  {author} {\bibinfo {author} {\bibfnamefont {I.}~\bibnamefont
  {Carusotto}}\ and\ \bibinfo {author} {\bibfnamefont {C.}~\bibnamefont
  {Ciuti}},\ }\href {\doibase 10.1103/RevModPhys.85.299} {\bibfield  {journal}
  {\bibinfo  {journal} {Rev. Mod. Phys.}\ }\textbf {\bibinfo {volume} {85}},\
  \bibinfo {pages} {299} (\bibinfo {year} {2013})}\BibitemShut {NoStop}%
\bibitem [{\citenamefont {Kasprzak}\ \emph {et~al.}(2006)\citenamefont
  {Kasprzak}, \citenamefont {Richard}, \citenamefont {Kundermann},
  \citenamefont {Baas}, \citenamefont {Jeambrun}, \citenamefont {Keeling},
  \citenamefont {Marchetti}, \citenamefont {Szymanska}, \citenamefont {Andre},
  \citenamefont {Staehli}, \citenamefont {Savona}, \citenamefont {Littlewood},
  \citenamefont {Deveaud},\ and\ \citenamefont {Dang}}]{Kasprzak_BEC}%
  \BibitemOpen
  \bibfield  {author} {\bibinfo {author} {\bibfnamefont {J.}~\bibnamefont
  {Kasprzak}}, \bibinfo {author} {\bibfnamefont {M.}~\bibnamefont {Richard}},
  \bibinfo {author} {\bibfnamefont {S.}~\bibnamefont {Kundermann}}, \bibinfo
  {author} {\bibfnamefont {A.}~\bibnamefont {Baas}}, \bibinfo {author}
  {\bibfnamefont {P.}~\bibnamefont {Jeambrun}}, \bibinfo {author}
  {\bibfnamefont {J.~M.~J.}\ \bibnamefont {Keeling}}, \bibinfo {author}
  {\bibfnamefont {F.~M.}\ \bibnamefont {Marchetti}}, \bibinfo {author}
  {\bibfnamefont {M.~H.}\ \bibnamefont {Szymanska}}, \bibinfo {author}
  {\bibfnamefont {R.}~\bibnamefont {Andre}}, \bibinfo {author} {\bibfnamefont
  {J.~L.}\ \bibnamefont {Staehli}}, \bibinfo {author} {\bibfnamefont
  {V.}~\bibnamefont {Savona}}, \bibinfo {author} {\bibfnamefont {P.~B.}\
  \bibnamefont {Littlewood}}, \bibinfo {author} {\bibfnamefont
  {B.}~\bibnamefont {Deveaud}}, \ and\ \bibinfo {author} {\bibfnamefont
  {L.~S.}\ \bibnamefont {Dang}},\ }\href {\doibase 10.1038/nature05131}
  {\bibfield  {journal} {\bibinfo  {journal} {Nature}\ }\textbf {\bibinfo
  {volume} {443}},\ \bibinfo {pages} {409} (\bibinfo {year}
  {2006})}\BibitemShut {NoStop}%
\bibitem [{\citenamefont {Amo}\ \emph {et~al.}(2009)\citenamefont {Amo},
  \citenamefont {Lefr{\`e}re}, \citenamefont {Pigeon}, \citenamefont {Adrados},
  \citenamefont {Ciuti}, \citenamefont {Carusotto}, \citenamefont {Houdr{\'e}},
  \citenamefont {Giacobino},\ and\ \citenamefont
  {Bramati}}]{Amo_Superfluidity}%
  \BibitemOpen
  \bibfield  {author} {\bibinfo {author} {\bibfnamefont {A.}~\bibnamefont
  {Amo}}, \bibinfo {author} {\bibfnamefont {J.}~\bibnamefont {Lefr{\`e}re}},
  \bibinfo {author} {\bibfnamefont {S.}~\bibnamefont {Pigeon}}, \bibinfo
  {author} {\bibfnamefont {C.}~\bibnamefont {Adrados}}, \bibinfo {author}
  {\bibfnamefont {C.}~\bibnamefont {Ciuti}}, \bibinfo {author} {\bibfnamefont
  {I.}~\bibnamefont {Carusotto}}, \bibinfo {author} {\bibfnamefont
  {R.}~\bibnamefont {Houdr{\'e}}}, \bibinfo {author} {\bibfnamefont
  {E.}~\bibnamefont {Giacobino}}, \ and\ \bibinfo {author} {\bibfnamefont
  {A.}~\bibnamefont {Bramati}},\ }\href {http://dx.doi.org/10.1038/nphys1364}
  {\bibfield  {journal} {\bibinfo  {journal} {Nature Physics}\ }\textbf
  {\bibinfo {volume} {5}},\ \bibinfo {pages} {805} (\bibinfo {year}
  {2009})}\BibitemShut {NoStop}%
\bibitem [{\citenamefont {Deng}\ \emph {et~al.}(2010)\citenamefont {Deng},
  \citenamefont {Haug},\ and\ \citenamefont {Yamamoto}}]{Yamamoto_RMP}%
  \BibitemOpen
  \bibfield  {author} {\bibinfo {author} {\bibfnamefont {H.}~\bibnamefont
  {Deng}}, \bibinfo {author} {\bibfnamefont {H.}~\bibnamefont {Haug}}, \ and\
  \bibinfo {author} {\bibfnamefont {Y.}~\bibnamefont {Yamamoto}},\ }\href
  {\doibase 10.1103/RevModPhys.82.1489} {\bibfield  {journal} {\bibinfo
  {journal} {Rev. Mod. Phys.}\ }\textbf {\bibinfo {volume} {82}},\ \bibinfo
  {pages} {1489} (\bibinfo {year} {2010})}\BibitemShut {NoStop}%
\bibitem [{\citenamefont {Ballarini}\ \emph {et~al.}(2013)\citenamefont
  {Ballarini}, \citenamefont {De~Giorgi}, \citenamefont {Cancellieri},
  \citenamefont {Houdr{\'e}}, \citenamefont {Giacobino}, \citenamefont
  {Cingolani}, \citenamefont {Bramati}, \citenamefont {Gigli},\ and\
  \citenamefont {Sanvitto}}]{Sanvitto_Transistor}%
  \BibitemOpen
  \bibfield  {author} {\bibinfo {author} {\bibfnamefont {D.}~\bibnamefont
  {Ballarini}}, \bibinfo {author} {\bibfnamefont {M.}~\bibnamefont
  {De~Giorgi}}, \bibinfo {author} {\bibfnamefont {E.}~\bibnamefont
  {Cancellieri}}, \bibinfo {author} {\bibfnamefont {R.}~\bibnamefont
  {Houdr{\'e}}}, \bibinfo {author} {\bibfnamefont {E.}~\bibnamefont
  {Giacobino}}, \bibinfo {author} {\bibfnamefont {R.}~\bibnamefont
  {Cingolani}}, \bibinfo {author} {\bibfnamefont {A.}~\bibnamefont {Bramati}},
  \bibinfo {author} {\bibfnamefont {G.}~\bibnamefont {Gigli}}, \ and\ \bibinfo
  {author} {\bibfnamefont {D.}~\bibnamefont {Sanvitto}},\ }\href
  {http://dx.doi.org/10.1038/ncomms2734} {\bibfield  {journal} {\bibinfo
  {journal} {Nature Communications}\ }\textbf {\bibinfo {volume} {4}},\
  \bibinfo {pages} {1778 EP } (\bibinfo {year} {2013})},\ \bibinfo {note}
  {article}\BibitemShut {NoStop}%
\bibitem [{\citenamefont {Gao}\ \emph {et~al.}(2012)\citenamefont {Gao},
  \citenamefont {Eldridge}, \citenamefont {Liew}, \citenamefont {Tsintzos},
  \citenamefont {Stavrinidis}, \citenamefont {Deligeorgis}, \citenamefont
  {Hatzopoulos},\ and\ \citenamefont {Savvidis}}]{Savvidis_TransistorSwitch}%
  \BibitemOpen
  \bibfield  {author} {\bibinfo {author} {\bibfnamefont {T.}~\bibnamefont
  {Gao}}, \bibinfo {author} {\bibfnamefont {P.~S.}\ \bibnamefont {Eldridge}},
  \bibinfo {author} {\bibfnamefont {T.~C.~H.}\ \bibnamefont {Liew}}, \bibinfo
  {author} {\bibfnamefont {S.~I.}\ \bibnamefont {Tsintzos}}, \bibinfo {author}
  {\bibfnamefont {G.}~\bibnamefont {Stavrinidis}}, \bibinfo {author}
  {\bibfnamefont {G.}~\bibnamefont {Deligeorgis}}, \bibinfo {author}
  {\bibfnamefont {Z.}~\bibnamefont {Hatzopoulos}}, \ and\ \bibinfo {author}
  {\bibfnamefont {P.~G.}\ \bibnamefont {Savvidis}},\ }\href {\doibase
  10.1103/PhysRevB.85.235102} {\bibfield  {journal} {\bibinfo  {journal} {Phys.
  Rev. B}\ }\textbf {\bibinfo {volume} {85}},\ \bibinfo {pages} {235102}
  (\bibinfo {year} {2012})}\BibitemShut {NoStop}%
\bibitem [{\citenamefont {Amo}\ \emph {et~al.}(2010)\citenamefont {Amo},
  \citenamefont {Liew}, \citenamefont {Adrados}, \citenamefont {Houdr{\'e}},
  \citenamefont {Giacobino}, \citenamefont {Kavokin},\ and\ \citenamefont
  {Bramati}}]{Bramati_SpinSwitches}%
  \BibitemOpen
  \bibfield  {author} {\bibinfo {author} {\bibfnamefont {A.}~\bibnamefont
  {Amo}}, \bibinfo {author} {\bibfnamefont {T.~C.~H.}\ \bibnamefont {Liew}},
  \bibinfo {author} {\bibfnamefont {C.}~\bibnamefont {Adrados}}, \bibinfo
  {author} {\bibfnamefont {R.}~\bibnamefont {Houdr{\'e}}}, \bibinfo {author}
  {\bibfnamefont {E.}~\bibnamefont {Giacobino}}, \bibinfo {author}
  {\bibfnamefont {A.~V.}\ \bibnamefont {Kavokin}}, \ and\ \bibinfo {author}
  {\bibfnamefont {A.}~\bibnamefont {Bramati}},\ }\href
  {http://dx.doi.org/10.1038/nphoton.2010.79} {\bibfield  {journal} {\bibinfo
  {journal} {Nature Photonics}\ }\textbf {\bibinfo {volume} {4}},\ \bibinfo
  {pages} {361 EP } (\bibinfo {year} {2010})}\BibitemShut {NoStop}%
\bibitem [{\citenamefont {{Espinosa-Ortega}}\ and\ \citenamefont
  {{Liew}}(2015)}]{Liew_Perceptrons}%
  \BibitemOpen
  \bibfield  {author} {\bibinfo {author} {\bibfnamefont {T.}~\bibnamefont
  {{Espinosa-Ortega}}}\ and\ \bibinfo {author} {\bibfnamefont {T.~C.~H.}\
  \bibnamefont {{Liew}}},\ }\href {\doibase 10.1103/PhysRevLett.114.118101}
  {\bibfield  {journal} {\bibinfo  {journal} {Physical Review Letters}\
  }\textbf {\bibinfo {volume} {114}},\ \bibinfo {eid} {118101} (\bibinfo {year}
  {2015})},\ \Eprint {http://arxiv.org/abs/1408.6949} {arXiv:1408.6949
  [cond-mat.dis-nn]} \BibitemShut {NoStop}%
\bibitem [{\citenamefont {Ohadi}\ \emph {et~al.}(2015)\citenamefont {Ohadi},
  \citenamefont {Dreismann}, \citenamefont {Rubo}, \citenamefont {Pinsker},
  \citenamefont {del Valle-Inclan~Redondo}, \citenamefont {Tsintzos},
  \citenamefont {Hatzopoulos}, \citenamefont {Savvidis},\ and\ \citenamefont
  {Baumberg}}]{Baumberg_SpinBifurcations}%
  \BibitemOpen
  \bibfield  {author} {\bibinfo {author} {\bibfnamefont {H.}~\bibnamefont
  {Ohadi}}, \bibinfo {author} {\bibfnamefont {A.}~\bibnamefont {Dreismann}},
  \bibinfo {author} {\bibfnamefont {Y.~G.}\ \bibnamefont {Rubo}}, \bibinfo
  {author} {\bibfnamefont {F.}~\bibnamefont {Pinsker}}, \bibinfo {author}
  {\bibfnamefont {Y.}~\bibnamefont {del Valle-Inclan~Redondo}}, \bibinfo
  {author} {\bibfnamefont {S.~I.}\ \bibnamefont {Tsintzos}}, \bibinfo {author}
  {\bibfnamefont {Z.}~\bibnamefont {Hatzopoulos}}, \bibinfo {author}
  {\bibfnamefont {P.~G.}\ \bibnamefont {Savvidis}}, \ and\ \bibinfo {author}
  {\bibfnamefont {J.~J.}\ \bibnamefont {Baumberg}},\ }\href {\doibase
  10.1103/PhysRevX.5.031002} {\bibfield  {journal} {\bibinfo  {journal} {Phys.
  Rev. X}\ }\textbf {\bibinfo {volume} {5}},\ \bibinfo {pages} {031002}
  (\bibinfo {year} {2015})}\BibitemShut {NoStop}%
\bibitem [{\citenamefont {LeCun}\ \emph {et~al.}(1998)\citenamefont {LeCun},
  \citenamefont {Bottou}, \citenamefont {Bengio},\ and\ \citenamefont
  {Haffner}}]{LeCun_MNIST}%
  \BibitemOpen
  \bibfield  {author} {\bibinfo {author} {\bibfnamefont {Y.}~\bibnamefont
  {LeCun}}, \bibinfo {author} {\bibfnamefont {L.}~\bibnamefont {Bottou}},
  \bibinfo {author} {\bibfnamefont {Y.}~\bibnamefont {Bengio}}, \ and\ \bibinfo
  {author} {\bibfnamefont {P.}~\bibnamefont {Haffner}},\ }\href {\doibase
  10.1109/5.726791} {\bibfield  {journal} {\bibinfo  {journal} {Proceedings of
  the IEEE}\ }\textbf {\bibinfo {volume} {86}},\ \bibinfo {pages} {2278}
  (\bibinfo {year} {1998})}\BibitemShut {NoStop}%
\bibitem [{\citenamefont {Bobrovska}\ \emph {et~al.}(2018)\citenamefont
  {Bobrovska}, \citenamefont {Matuszewski}, \citenamefont {Daskalakis},
  \citenamefont {Maier},\ and\ \citenamefont {K{\'e}na-Cohen}}]{Bobrovska_DI}%
  \BibitemOpen
  \bibfield  {author} {\bibinfo {author} {\bibfnamefont {N.}~\bibnamefont
  {Bobrovska}}, \bibinfo {author} {\bibfnamefont {M.}~\bibnamefont
  {Matuszewski}}, \bibinfo {author} {\bibfnamefont {K.~S.}\ \bibnamefont
  {Daskalakis}}, \bibinfo {author} {\bibfnamefont {S.~A.}\ \bibnamefont
  {Maier}}, \ and\ \bibinfo {author} {\bibfnamefont {S.}~\bibnamefont
  {K{\'e}na-Cohen}},\ }\href {\doibase 10.1021/acsphotonics.7b00283} {\bibfield
   {journal} {\bibinfo  {journal} {ACS Photonics}\ }\textbf {\bibinfo {volume}
  {5}},\ \bibinfo {pages} {111} (\bibinfo {year} {2018})}\BibitemShut {NoStop}%
\bibitem [{\citenamefont {Nardin}\ \emph {et~al.}(2011)\citenamefont {Nardin},
  \citenamefont {Grosso}, \citenamefont {L{\'e}ger}, \citenamefont {Pietka},
  \citenamefont {Morier-Genoud},\ and\ \citenamefont
  {Deveaud-Pl{\'e}dran}}]{Deveaud_HydrodynamicVortices}%
  \BibitemOpen
  \bibfield  {author} {\bibinfo {author} {\bibfnamefont {G.}~\bibnamefont
  {Nardin}}, \bibinfo {author} {\bibfnamefont {G.}~\bibnamefont {Grosso}},
  \bibinfo {author} {\bibfnamefont {Y.}~\bibnamefont {L{\'e}ger}}, \bibinfo
  {author} {\bibfnamefont {B.}~\bibnamefont {Pietka}}, \bibinfo {author}
  {\bibfnamefont {F.}~\bibnamefont {Morier-Genoud}}, \ and\ \bibinfo {author}
  {\bibfnamefont {B.}~\bibnamefont {Deveaud-Pl{\'e}dran}},\ }\href
  {http://dx.doi.org/10.1038/nphys1959} {\bibfield  {journal} {\bibinfo
  {journal} {Nature Physics}\ }\textbf {\bibinfo {volume} {7}},\ \bibinfo
  {pages} {635 EP } (\bibinfo {year} {2011})},\ \bibinfo {note}
  {article}\BibitemShut {NoStop}%
\bibitem [{\citenamefont {Abbarchi}\ \emph {et~al.}(2013)\citenamefont
  {Abbarchi}, \citenamefont {Amo}, \citenamefont {Sala}, \citenamefont
  {Solnyshkov}, \citenamefont {Flayac}, \citenamefont {Ferrier}, \citenamefont
  {Sagnes}, \citenamefont {Galopin}, \citenamefont {Lema{\^i}tre},
  \citenamefont {Malpuech},\ and\ \citenamefont {Bloch}}]{Bloch_MQST}%
  \BibitemOpen
  \bibfield  {author} {\bibinfo {author} {\bibfnamefont {M.}~\bibnamefont
  {Abbarchi}}, \bibinfo {author} {\bibfnamefont {A.}~\bibnamefont {Amo}},
  \bibinfo {author} {\bibfnamefont {V.~G.}\ \bibnamefont {Sala}}, \bibinfo
  {author} {\bibfnamefont {D.~D.}\ \bibnamefont {Solnyshkov}}, \bibinfo
  {author} {\bibfnamefont {H.}~\bibnamefont {Flayac}}, \bibinfo {author}
  {\bibfnamefont {L.}~\bibnamefont {Ferrier}}, \bibinfo {author} {\bibfnamefont
  {I.}~\bibnamefont {Sagnes}}, \bibinfo {author} {\bibfnamefont
  {E.}~\bibnamefont {Galopin}}, \bibinfo {author} {\bibfnamefont
  {A.}~\bibnamefont {Lema{\^i}tre}}, \bibinfo {author} {\bibfnamefont
  {G.}~\bibnamefont {Malpuech}}, \ and\ \bibinfo {author} {\bibfnamefont
  {J.}~\bibnamefont {Bloch}},\ }\href {http://dx.doi.org/10.1038/nphys2609}
  {\bibfield  {journal} {\bibinfo  {journal} {Nature Physics}\ }\textbf
  {\bibinfo {volume} {9}},\ \bibinfo {pages} {275 EP } (\bibinfo {year}
  {2013})}\BibitemShut {NoStop}%
\bibitem [{\citenamefont {Cilibrizzi}\ \emph {et~al.}(2016)\citenamefont
  {Cilibrizzi}, \citenamefont {Sigurdsson}, \citenamefont {Liew}, \citenamefont
  {Ohadi}, \citenamefont {Askitopoulos}, \citenamefont {Brodbeck},
  \citenamefont {Schneider}, \citenamefont {Shelykh}, \citenamefont
  {H\"ofling}, \citenamefont {Ruostekoski},\ and\ \citenamefont
  {Lagoudakis}}]{Lagoudakis_Skyrmion}%
  \BibitemOpen
  \bibfield  {author} {\bibinfo {author} {\bibfnamefont {P.}~\bibnamefont
  {Cilibrizzi}}, \bibinfo {author} {\bibfnamefont {H.}~\bibnamefont
  {Sigurdsson}}, \bibinfo {author} {\bibfnamefont {T.~C.~H.}\ \bibnamefont
  {Liew}}, \bibinfo {author} {\bibfnamefont {H.}~\bibnamefont {Ohadi}},
  \bibinfo {author} {\bibfnamefont {A.}~\bibnamefont {Askitopoulos}}, \bibinfo
  {author} {\bibfnamefont {S.}~\bibnamefont {Brodbeck}}, \bibinfo {author}
  {\bibfnamefont {C.}~\bibnamefont {Schneider}}, \bibinfo {author}
  {\bibfnamefont {I.~A.}\ \bibnamefont {Shelykh}}, \bibinfo {author}
  {\bibfnamefont {S.}~\bibnamefont {H\"ofling}}, \bibinfo {author}
  {\bibfnamefont {J.}~\bibnamefont {Ruostekoski}}, \ and\ \bibinfo {author}
  {\bibfnamefont {P.}~\bibnamefont {Lagoudakis}},\ }\href {\doibase
  10.1103/PhysRevB.94.045315} {\bibfield  {journal} {\bibinfo  {journal} {Phys.
  Rev. B}\ }\textbf {\bibinfo {volume} {94}},\ \bibinfo {pages} {045315}
  (\bibinfo {year} {2016})}\BibitemShut {NoStop}%
\bibitem [{\citenamefont {Pieczarka}\ \emph {et~al.}(2017)\citenamefont
  {Pieczarka}, \citenamefont {Syperek}, \citenamefont {Dusanowski},
  \citenamefont {Opala}, \citenamefont {Langer}, \citenamefont {Schneider},
  \citenamefont {H{\"o}fling},\ and\ \citenamefont
  {Sek}}]{Pieczarka_RelaxationOscillations}%
  \BibitemOpen
  \bibfield  {author} {\bibinfo {author} {\bibfnamefont {M.}~\bibnamefont
  {Pieczarka}}, \bibinfo {author} {\bibfnamefont {M.}~\bibnamefont {Syperek}},
  \bibinfo {author} {\bibfnamefont {L.}~\bibnamefont {Dusanowski}}, \bibinfo
  {author} {\bibfnamefont {A.}~\bibnamefont {Opala}}, \bibinfo {author}
  {\bibfnamefont {F.}~\bibnamefont {Langer}}, \bibinfo {author} {\bibfnamefont
  {C.}~\bibnamefont {Schneider}}, \bibinfo {author} {\bibfnamefont
  {S.}~\bibnamefont {H{\"o}fling}}, \ and\ \bibinfo {author} {\bibfnamefont
  {G.}~\bibnamefont {Sek}},\ }\href {\doibase 10.1038/s41598-017-07470-8}
  {\bibfield  {journal} {\bibinfo  {journal} {Scientific Reports}\ }\textbf
  {\bibinfo {volume} {7}},\ \bibinfo {pages} {7094} (\bibinfo {year}
  {2017})}\BibitemShut {NoStop}%
\bibitem [{\citenamefont {{Wang}}\ \emph {et~al.}(2016)\citenamefont {{Wang}},
  \citenamefont {{Li}},\ and\ \citenamefont {{Li}}}]{Wang_LSM_FPGA}%
  \BibitemOpen
  \bibfield  {author} {\bibinfo {author} {\bibfnamefont {Q.}~\bibnamefont
  {{Wang}}}, \bibinfo {author} {\bibfnamefont {Y.}~\bibnamefont {{Li}}}, \ and\
  \bibinfo {author} {\bibfnamefont {P.}~\bibnamefont {{Li}}},\ }in\ \href
  {\doibase 10.1109/ISCAS.2016.7527245} {\emph {\bibinfo {booktitle} {2016 IEEE
  International Symposium on Circuits and Systems (ISCAS)}}}\ (\bibinfo {year}
  {2016})\ pp.\ \bibinfo {pages} {361--364}\BibitemShut {NoStop}%
\bibitem [{\citenamefont {Mili{\'{c}}evi{\'{c}}}\ \emph
  {et~al.}(2015)\citenamefont {Mili{\'{c}}evi{\'{c}}}, \citenamefont {Ozawa},
  \citenamefont {Andreakou}, \citenamefont {Carusotto}, \citenamefont
  {Jacqmin}, \citenamefont {Galopin}, \citenamefont {Lema{\^{\i}}tre},
  \citenamefont {Gratiet}, \citenamefont {Sagnes}, \citenamefont {Bloch},\ and\
  \citenamefont {Amo}}]{Bloch_EdgeStates}%
  \BibitemOpen
  \bibfield  {author} {\bibinfo {author} {\bibfnamefont {M.}~\bibnamefont
  {Mili{\'{c}}evi{\'{c}}}}, \bibinfo {author} {\bibfnamefont {T.}~\bibnamefont
  {Ozawa}}, \bibinfo {author} {\bibfnamefont {P.}~\bibnamefont {Andreakou}},
  \bibinfo {author} {\bibfnamefont {I.}~\bibnamefont {Carusotto}}, \bibinfo
  {author} {\bibfnamefont {T.}~\bibnamefont {Jacqmin}}, \bibinfo {author}
  {\bibfnamefont {E.}~\bibnamefont {Galopin}}, \bibinfo {author} {\bibfnamefont
  {A.}~\bibnamefont {Lema{\^{\i}}tre}}, \bibinfo {author} {\bibfnamefont
  {L.~L.}\ \bibnamefont {Gratiet}}, \bibinfo {author} {\bibfnamefont
  {I.}~\bibnamefont {Sagnes}}, \bibinfo {author} {\bibfnamefont
  {J.}~\bibnamefont {Bloch}}, \ and\ \bibinfo {author} {\bibfnamefont
  {A.}~\bibnamefont {Amo}},\ }\href {\doibase 10.1088/2053-1583/2/3/034012}
  {\bibfield  {journal} {\bibinfo  {journal} {2D Materials}\ }\textbf {\bibinfo
  {volume} {2}},\ \bibinfo {pages} {034012} (\bibinfo {year}
  {2015})}\BibitemShut {NoStop}%
\bibitem [{\citenamefont {Merolla}\ \emph {et~al.}(2014)\citenamefont
  {Merolla}, \citenamefont {Arthur}, \citenamefont {Alvarez-Icaza},
  \citenamefont {Cassidy}, \citenamefont {Sawada}, \citenamefont {Akopyan},
  \citenamefont {Jackson}, \citenamefont {Imam}, \citenamefont {Guo},
  \citenamefont {Nakamura}, \citenamefont {Brezzo}, \citenamefont {Vo},
  \citenamefont {Esser}, \citenamefont {Appuswamy}, \citenamefont {Taba},
  \citenamefont {Amir}, \citenamefont {Flickner}, \citenamefont {Risk},
  \citenamefont {Manohar},\ and\ \citenamefont
  {Modha}}]{Merolla_MillionSpkingNeurons}%
  \BibitemOpen
  \bibfield  {author} {\bibinfo {author} {\bibfnamefont {P.~A.}\ \bibnamefont
  {Merolla}}, \bibinfo {author} {\bibfnamefont {J.~V.}\ \bibnamefont {Arthur}},
  \bibinfo {author} {\bibfnamefont {R.}~\bibnamefont {Alvarez-Icaza}}, \bibinfo
  {author} {\bibfnamefont {A.~S.}\ \bibnamefont {Cassidy}}, \bibinfo {author}
  {\bibfnamefont {J.}~\bibnamefont {Sawada}}, \bibinfo {author} {\bibfnamefont
  {F.}~\bibnamefont {Akopyan}}, \bibinfo {author} {\bibfnamefont {B.~L.}\
  \bibnamefont {Jackson}}, \bibinfo {author} {\bibfnamefont {N.}~\bibnamefont
  {Imam}}, \bibinfo {author} {\bibfnamefont {C.}~\bibnamefont {Guo}}, \bibinfo
  {author} {\bibfnamefont {Y.}~\bibnamefont {Nakamura}}, \bibinfo {author}
  {\bibfnamefont {B.}~\bibnamefont {Brezzo}}, \bibinfo {author} {\bibfnamefont
  {I.}~\bibnamefont {Vo}}, \bibinfo {author} {\bibfnamefont {S.~K.}\
  \bibnamefont {Esser}}, \bibinfo {author} {\bibfnamefont {R.}~\bibnamefont
  {Appuswamy}}, \bibinfo {author} {\bibfnamefont {B.}~\bibnamefont {Taba}},
  \bibinfo {author} {\bibfnamefont {A.}~\bibnamefont {Amir}}, \bibinfo {author}
  {\bibfnamefont {M.~D.}\ \bibnamefont {Flickner}}, \bibinfo {author}
  {\bibfnamefont {W.~P.}\ \bibnamefont {Risk}}, \bibinfo {author}
  {\bibfnamefont {R.}~\bibnamefont {Manohar}}, \ and\ \bibinfo {author}
  {\bibfnamefont {D.~S.}\ \bibnamefont {Modha}},\ }\href {\doibase
  10.1126/science.1254642} {\bibfield  {journal} {\bibinfo  {journal}
  {Science}\ }\textbf {\bibinfo {volume} {345}},\ \bibinfo {pages} {668}
  (\bibinfo {year} {2014})},\ \Eprint
  {http://arxiv.org/abs/http://science.sciencemag.org/content/345/6197/668.full.pdf}
  {http://science.sciencemag.org/content/345/6197/668.full.pdf} \BibitemShut
  {NoStop}%
\bibitem [{\citenamefont {Li}\ \emph {et~al.}(2012)\citenamefont {Li},
  \citenamefont {Han},\ and\ \citenamefont {Wang}}]{Li2012}%
  \BibitemOpen
  \bibfield  {author} {\bibinfo {author} {\bibfnamefont {D.}~\bibnamefont
  {Li}}, \bibinfo {author} {\bibfnamefont {M.}~\bibnamefont {Han}}, \ and\
  \bibinfo {author} {\bibfnamefont {J.}~\bibnamefont {Wang}},\ }\href {\doibase
  10.1109/TNNLS.2012.2188414} {\bibfield  {journal} {\bibinfo  {journal} {IEEE
  Transactions on Neural Networks and Learning Systems}\ }\textbf {\bibinfo
  {volume} {23}},\ \bibinfo {pages} {787} (\bibinfo {year} {2012})}\BibitemShut
  {NoStop}%
\bibitem [{\citenamefont {Lyon}(1982)}]{Lyon1982}%
  \BibitemOpen
  \bibfield  {author} {\bibinfo {author} {\bibfnamefont {R.}~\bibnamefont
  {Lyon}},\ }in\ \href {\doibase 10.1109/ICASSP.1982.1171644} {\emph {\bibinfo
  {booktitle} {ICASSP '82. IEEE International Conference on Acoustics, Speech,
  and Signal Processing}}},\ Vol.~\bibinfo {volume} {7}\ (\bibinfo {year}
  {1982})\ pp.\ \bibinfo {pages} {1282--1285}\BibitemShut {NoStop}%
\bibitem [{\citenamefont {Verstraeten}\ \emph {et~al.}(2005)\citenamefont
  {Verstraeten}, \citenamefont {Schrauwen}, \citenamefont {Stroobandt},\ and\
  \citenamefont {Campenhout}}]{Verstraeten2005}%
  \BibitemOpen
  \bibfield  {author} {\bibinfo {author} {\bibfnamefont {D.}~\bibnamefont
  {Verstraeten}}, \bibinfo {author} {\bibfnamefont {B.}~\bibnamefont
  {Schrauwen}}, \bibinfo {author} {\bibfnamefont {D.}~\bibnamefont
  {Stroobandt}}, \ and\ \bibinfo {author} {\bibfnamefont {J.~V.}\ \bibnamefont
  {Campenhout}},\ }\href {\doibase https://doi.org/10.1016/j.ipl.2005.05.019}
  {\bibfield  {journal} {\bibinfo  {journal} {Information Processing Letters}\
  }\textbf {\bibinfo {volume} {95}},\ \bibinfo {pages} {521 } (\bibinfo {year}
  {2005})},\ \bibinfo {note} {applications of Spiking Neural
  Networks}\BibitemShut {NoStop}%
\bibitem [{\citenamefont {Duan}\ \emph {et~al.}(2000)\citenamefont {Duan},
  \citenamefont {Giedke}, \citenamefont {Cirac},\ and\ \citenamefont
  {Zoller}}]{Duan2000}%
  \BibitemOpen
  \bibfield  {author} {\bibinfo {author} {\bibfnamefont {L.-M.}\ \bibnamefont
  {Duan}}, \bibinfo {author} {\bibfnamefont {G.}~\bibnamefont {Giedke}},
  \bibinfo {author} {\bibfnamefont {J.~I.}\ \bibnamefont {Cirac}}, \ and\
  \bibinfo {author} {\bibfnamefont {P.}~\bibnamefont {Zoller}},\ }\href
  {\doibase 10.1103/PhysRevLett.84.2722} {\bibfield  {journal} {\bibinfo
  {journal} {Phys. Rev. Lett.}\ }\textbf {\bibinfo {volume} {84}},\ \bibinfo
  {pages} {2722} (\bibinfo {year} {2000})}\BibitemShut {NoStop}%
\bibitem [{\citenamefont {Simon}(2000)}]{Simon2000}%
  \BibitemOpen
  \bibfield  {author} {\bibinfo {author} {\bibfnamefont {R.}~\bibnamefont
  {Simon}},\ }\href {\doibase 10.1103/PhysRevLett.84.2726} {\bibfield
  {journal} {\bibinfo  {journal} {Phys. Rev. Lett.}\ }\textbf {\bibinfo
  {volume} {84}},\ \bibinfo {pages} {2726} (\bibinfo {year}
  {2000})}\BibitemShut {NoStop}%
\end{thebibliography}%
%
\end{document}